\documentclass[
reprint,
amsmath,amssymb,
aps,
]{revtex4-2}

\usepackage{amsmath,gensymb,textcomp,bm,dcolumn,eurosym,array,tabu,multirow,nicefrac,color,subfigure,graphicx,upgreek}
\usepackage[colorlinks, linkcolor=blue, citecolor=blue, urlcolor=blue, breaklinks=true]{hyperref}

\usepackage{txfonts,times} 

\newcommand{\bra}[1]{\langle #1 \rvert}
\newcommand{\ket}[1]{\lvert #1 \rangle}

\usepackage[subsectionbib]{bibunits}
\usepackage{graphicx}
\usepackage{dcolumn}
\usepackage{bm,MnSymbol}
\usepackage{units}
\usepackage{psfrag}
\usepackage{float}
\usepackage{color}

\begin{document}
	\preprint{APS/123-QED}
	\title{Ultrasensitive Transverse Deflection Measurement with Two-photon Interference}
	\author{Chaojie Wang}
    \author{Yuning Zhang}
	\author{Yuanyuan Chen}
	\email{chenyy@xmu.edu.cn}
	\author{Lixiang Chen}
	\email{chenlx@xmu.edu.cn}
	\affiliation{Department of Physics, Xiamen University, Xiamen 361005, China}%
	\begin{abstract}
Hong-Ou-Mandel (HOM) interference is an intrinsic quantum phenomena that goes beyond the possibilities of classical physics, and enables various applications in quantum metrology. While the timing resolution of HOM-based sensor is generally investigated, the ultimate quantum sensitivity in the estimation of transverse deflection or displacement between paired photons interfering at a balanced beam splitter has been explored relatively little. Here, we present an experimental demonstration of a spatial HOM interferometry for measuring the transverse deflection of an optical beam by using transverse momentum sampling measurements. This feasible scheme suffices to achieve great precision with comparatively little technological effort, which circumvents the stringent requirement in direct imaging resolution at the diffraction limit. We can adaptively determine optimum working points using a Fisher information analysis, and demonstrate an optimized spatial HOM interferometry according to practical applications. These results may significantly facilitate the use of quantum interference for high precision spatial sensing, and pave the way to more complex quantum imaging techniques like nanoscopic microscopy.
   \end{abstract}
	\maketitle
\section{Introduction}
The precise estimation of a physical quantity is essential in various research areas and practical applications. While optical interferometry can provide sensitive metrology capabilities, the best phase sensitivity achievable using n photons is the shot-noise limit $\Delta\phi=1/\sqrt{n}$ \cite{giovannetti2011advances,slussarenko2017unconditional}. Moreover, the environmental instability also poses a fundamental limit in the repeatability and measurement accuracy. The exploitation of quantum interference promises to enhance sensing technologies beyond the possibilities of classical physics \cite{giovannetti2004quantum,wolfgramm2013entanglement,holland1993interferometric,resch2007time,mitchell2004super,nagata2007beating}. Hong-Ou-Mandel (HOM) interference is a prototypical example of such a quantum phenomena, which states the fact that identical photons that arrive simultaneously on different input ports of a beam splitter would bunch into a common output port \cite{hong1987measurement}. On the other hand, for paired photons that are nonidentical in times, polarization or central frequencies, the ``bunching'' probability is directly related to the photons' level of indistinguishability or its degree of purity. Thus, HOM interferometry enables a wide range of high precision measurement tasks, ranging from measuring optical delays between different paths \cite{lyons2018attoseond,chen2019hong}, quantum optical coherence tomography \cite{nasr2003demonstration,Ibarra-Borja:20,chen2023spectrally}, and quantum enhanced imaging techniques \cite{ndagano2022quantum,devaux2020imaging}, to name but a few. 

It has been well known that inner variables resolved optical interference can circumvent the requirement of large overlap in the photonic wave packets \cite{devid1991oprical,bouma2022optical}. As a direct result, this technique enables the observation of beating signals, which reveals the interference pattern as a period oscillations within a coherent temporal envelope. Hence, a biphoton beat note has been proposed that suffices to achieve great prevision with a larger scaling that is relevant to the difference frequency of discrete color entangled states \cite{Legero2004quantum,ramelow2009discrete,chen2019hong,tamma2015multiboson}. Since the period in the quantum beating pattern is inversely proportional to the difference in the colors of incident photons, the frequency-resolving detection of two delayed photons that are impinged from opposite input ports of the beam splitter can provide precise determination of this temporal delay. This has been known as a quantum version of spectral-domain optical coherence tomography \cite{marchand2021soliton,Leitgeb2003optical}. With the assistance of holographic grating and single-photon cameras, quantum optical coherence tomography can speedup the information extraction, and still preserve its sensitivity in the case of nonoverlapping temporal packets \cite{chen2023spectrally,rank2021toward,Song2021review}. 

In addition to energy-time domain, two photon interference has also been performed in the spatial domain by varying transversal properties of the two photonic wave packets. Spatial HOM interferometry so far has mostly been exploited to access the spatial coherence of highly entangled photons produced by spontaneous parametric down conversion process \cite{ou1989further,kim2006spatial,lee2006spatial,devaux2020imaging}. While the spatiotemporal HOM interference has been imaged by measuring momentum spatial coincidences, the metrological potential of spatial two-photon interference for high precision sensing application has been explored relatively little. In particular, recent development of high-precision nanoscopic techniques like single photon cameras makes this spatial HOM interferometry even more compelling \cite{hanne2015sted,bruschini2019single,urban2021quantum}. Recently, a quantum interference technique based on spatially resolved sampling measurements has been designed and theoretically proved to be an optimal metrological scheme for estimating the transverse separation of their wave packets \cite{triggiani2024estimation}. While this elaborate scheme can be done by employing two cameras that spatially sample over all the possible two-photon interference events in the far field to resolve the difference in the transverse momenta, an experimental implementation with comparatively little technological effort is of great significance to practical applications.

Here, we experimentally demonstrate a ultrasensitive spatial HOM interferometry by using superpositions of two well-separated yet coherent spatial modes and coincidence detection on the biphoton spatial correlations. To introduce a specific parameter in spatial mode that to be estimated, a transverse deflection of one optical beam that is incident on the HOM interferometer is used in our experiment. The precise measurement of a tiny transverse deflection holds great promises for various sensing schemes like the localization and tracking of biological samples, high-precision refractometry and astrophysical bodies localizations. Building on the measurement and estimation strategy by analyzing the Fisher information, we explore the sensitivity limits as a function of the transverse deflection as imposed by the Cramér-Rao bound \cite{helstrom1969quantum,FUJIWARA1995119}, and find that the precision with which the deflection can be measured is mainly determined by the transverse-momentum distributions. Furthermore, instead of employing two cameras that spatially sample over all the possible two-photon interference events in the far field, we merely use a linear scan of transverse momenta that is triggered by the detection of its paired photon to resolve the difference in the transverse momenta of the two detected photons, which confirms that suitable spatially resolved sampling measurements are readily obtained with comparatively little technological efforts.

These results show that quantum interference of unconventional spatial states on a beam splitter provides a concise yet efficient way of enhancing the spatial resolution in HOM-based sensors and may also indicate a new direction towards fully harnessing HOM interference in quantum sensing and quantum information processing \cite{laibacher2015from,tamma2016multi,ober2004localization,lelek2021single,salazar2015tunable}.

\section{Theory}

\begin{figure}
	\centering
	\includegraphics[width=0.85\linewidth]{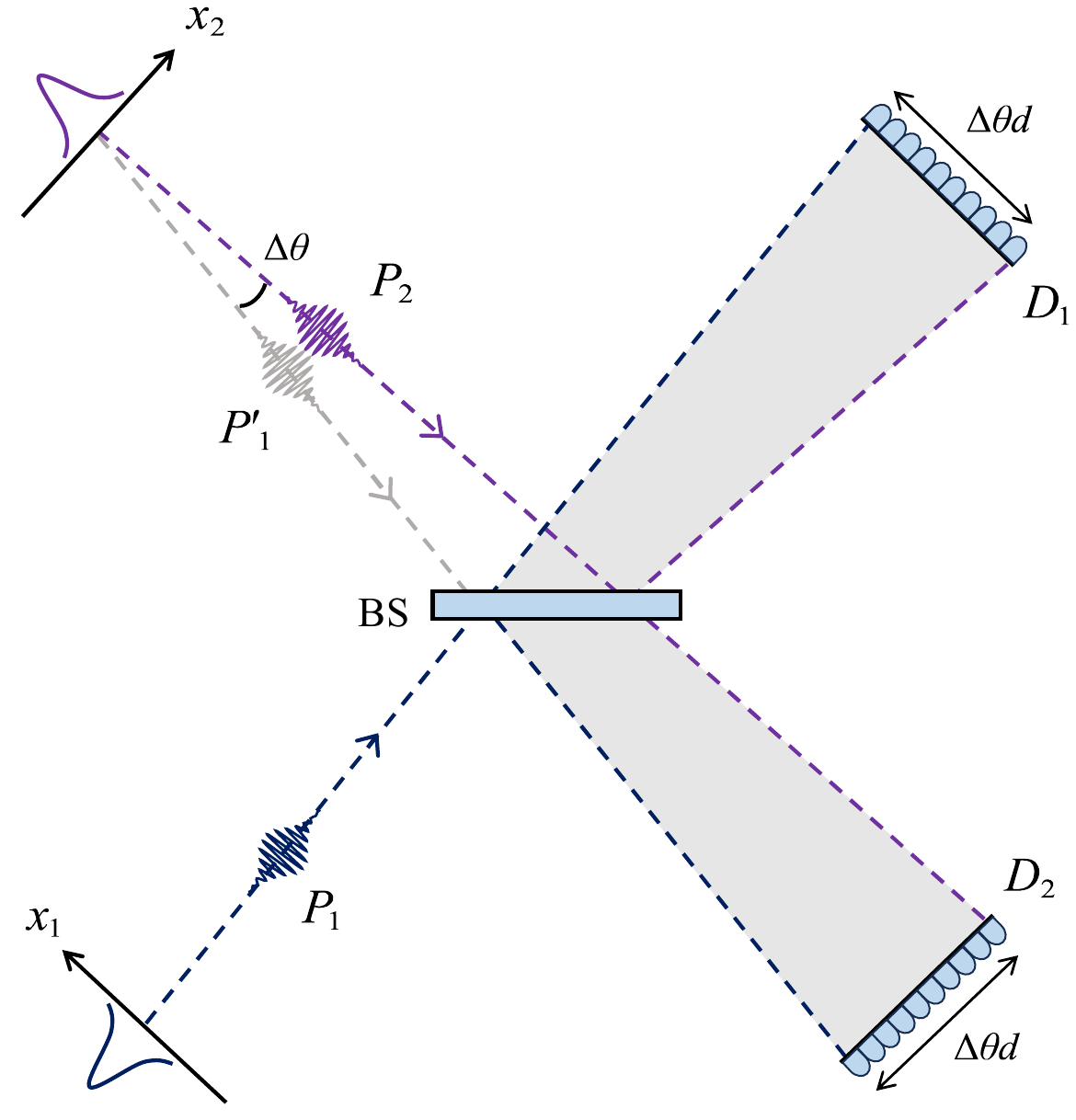}
	\caption{Schematic for the precise measurement of tiny transverse deflection. Two beams of light are incident on a balanced beam splitter (BS) from the opposite input ports, wherein a relative transverse deflection $\Delta \theta$ is introduced between the probe beam $P_2$ and the symmetric image $P_1^\prime$ of the reference beam $P_1$. Then, they are detected by two single photon detectors ($D_1$ and $D_2$) in the far-field regime, and the transverse displacement between the paired photons in the  detection screen is labeled as $\Delta \theta d $. }
    \label{fig1}
\end{figure}

Let us consider the generic task of measuring the transverse deflections of an optical beam as shown in Fig. \ref{fig1}. Assuming that the transverse position distribution of single photon is described by $\psi(x)$, thus a quantum system composed of two photons can be expressed as 
\begin{equation}
	\ket{\Psi}=\int d x_1\psi_1(x_1)\hat{a}_1^\dagger(x_1) \ket{0}_1\otimes\int d x_2\psi_2(x_2)\hat{a}_2^\dagger(x_2) \ket{0}_2,
\end{equation}
where the subscript S=1,2 denote two independent single photons, $\hat{a}_1^\dagger(x_1) $ and $\hat{a}_2^\dagger(x_2)$ are the bosonic creation operators associated with the first and second input mode of the beam splitter at transverse positions $x_1$ and $x_2$, and $\ket{0}$ represents the vacuum state. An unknown transverse deflection $\Delta\theta$ would transform the probe state $\ket{\Psi}$ to $\ket{\Psi(\Delta\theta)}$ upon interaction with the physical system. In our work, this interaction process can be demonstrated by an unitary evolution as $\hat{U}(\theta_1,\theta_2)=\exp(-i\theta_1d\hat{k}_1)\otimes \exp(-i\theta_2d\hat{k}_2)$, where $\theta_1-\theta_2=\Delta \theta$, $d$ is the distance between the source and the detector, $\hat{k}_1$ and $\hat{k}_2$ denote transverse momenta that are conjugate variables to the photon positions. The transformed state is then subjected to a particular measurement strategy to obtain an estimator of $\Delta\theta$, a fundamental limit for the precision of estimation is stated as \cite{helstrom1969quantum,FUJIWARA1995119}
\begin{equation}
\begin{split}
\text{Var}[\widetilde{\Delta \theta}]\geqslant\frac{1}{NH(\Delta \theta)},
\end{split}
\end{equation}
where
\begin{equation}
\begin{split}
H(\Delta \theta)=\langle\frac{\partial\Psi(\Delta \theta)}{\partial\Delta \theta}\ket{\frac{\partial\Psi(\Delta \theta)}{\partial\Delta \theta}}-|\langle\Psi(\Delta \theta)\ket{\frac{\partial\Psi(\Delta \theta)}{\partial\Delta \theta}}|^2,
\end{split}
\end{equation}
and $N$ is the number of independent sampling measurements, $H(\Delta \theta)$ is the quantum Fisher information. This statement is known as Quantum Cramér-Rao bound, which indicates that the measurement precision is ultimately limited by the probe state and interaction process, but independent of any ingenious measurement strategy. Namely, the appropriate choice of the probe state is of the utmost importance \cite{pirandola2018advances,chen2019hong}. For this work, the resultant quantum Fisher information is $H(\Delta \theta)=2\sigma_k^2d^2$, where $\sigma_k$ is the standard deviation of the transverse-momentum distribution of single photons (see supplementary for details).

\begin{figure*}
	\centering
\includegraphics[width=1\linewidth]{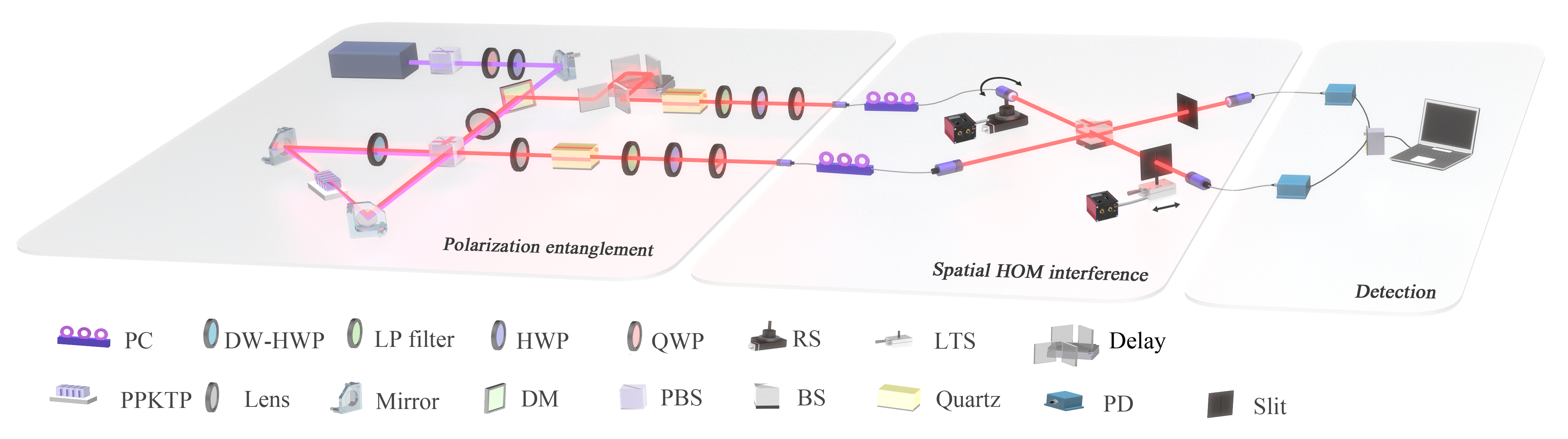}
\caption{ Experimental demonstration of the transverse deflection measurement with two-photon interference. PC, polarization controller; DW-HWP, dual-wavelength half-wave plate; LP filter, long-pass filter; HWP, half-wave plate; QWP, quarter wave plate; RS, rotation stage; LTS, linear translation stage; PPKTP, type-II periodically poled potassium titanyl phosphate crystal; DM, dichroic mirror; PBS, polarizing beam splitter; BS, balanced beam splitter; PD, single-photon avalanche detector.}
    \label{fig2}
\end{figure*}
While the measurement precision is inherent to the particular choice of a probe state and the interaction process, we still need an optimal measurement strategy to realize this potential benefit, namely one that allows us to saturate $H(\Delta \theta)$. We can accomplish this task with coincidence detection in the output ports of a balanced beam splitter, which constitutes a HOM interferometry. While the conventional HOM interferometry has the stringent requirement of spatial overlap, a transverse deflection angle $\Delta \theta$ can be introduced by slightly tilt one photonic beam. Since the detectors are positioned far from the sources with a distance $d$, a resultant spatial separation of $\Delta \theta d$ makes the paired photons are partly distinguishable. For specific positions on two detectors as $y_1$ and $y_2$, their corresponding unitary operator of the beam splitter and the detection operators on different output modes can be written as
\begin{equation}
\begin{aligned}
\hat{E}_{1}^{(+)}(y_1)&=\frac{\int{d}x_1\hat{a}_1(x_1)e^{-ik_1\left(x_1-\Delta\theta d\right)}+\int{d}x_2\hat{a}_2(x_2)e^{-ik_1x_2}}{\sqrt{4\pi}}, \\
 \hat{E}_{2}^{(+)}(y_2)&=\frac{\int{d}x_1\hat{a}_1(x_1)e^{-ik_2\left(x_1-\Delta\theta d\right)}+\int{d}x_2\hat{a}_2(x_2)e^{-ik_2x_2}}{\sqrt{4\pi}},
\end{aligned}	
\end{equation}
where $k_1=k _0y_1/d$ and $k_2=k_0y_2/d$ represent the transverse momentum of this paired photons, $k_0$ is the wave number. As a direct result, the normalized coincidence detection probability identified by single-photon detectors at distinct spatial modes reads
\begin{equation}
\begin{aligned}
	P_c(\Delta k,\Delta \theta )&=\bra{\Psi} \hat{E}_{1}^{(-)}(y_1) \hat{E}_{2}^{(-}(y_2)\hat{E}_{1}^{(+)}(y_1) \hat{E}_{2}^{(+)}(y_2)\ket{\Psi}\\
    &=\frac{1}{2}C(\Delta k)[1-\cos(\Delta k\Delta \theta d)]
    \label{eq: probability}
    \end{aligned}
	\end{equation}
where $\Delta k=k_1-k_2$, $C(\Delta k)=\exp (-\Delta k^2/4\sigma_k^2)/\sqrt{4\pi\sigma_k^2}$ is the modulo squared  $|\varphi(k)|^2$ of the Fourier transform of $\psi(x)$, which indicates the beats envelope that is determined by the transverse momentum distribution \cite{triggiani2024estimation,kim2006spatial,lee2006spatial}.

In the case of a real HOM interferometer that is subject to photon loss $\gamma$ and and imperfect experimental visibility $\nu$, there are three possible measurement outcomes; either both photons are detected as coincidence event, one photon is detected, or no photon is detected. The corresponding probability distributions become
\begin{equation}
	\begin{aligned}
		&P_0=\gamma ^2C(\Delta k)\\
		&P_1=\frac{1}{2}(1-\gamma )^2C(\Delta k)\left[ \frac{1+3\gamma}{1-\gamma}+\nu\cos(\Delta k\Delta \theta d) \right] \\
		&P_2=\frac{1}{2}(1-\gamma )^2C(\Delta k)\left[ 1-\nu \cos(\Delta k\Delta \theta d) \right] 
	\end{aligned}
	\end{equation}
where subscripts 0, 1, and 2 denote the number of detectors that click, corresponding to total loss, bunching, and anti-bunching, respectively. The outcome probabilities in this measurement can now be used to construct an estimator for the value of $\Delta \theta$ \cite{lyons2018attoseond}. The estimator $\widetilde{\Delta \theta}$ is a function of the experimental data that can be used to infer the value of the unknown transverse deflections using a specific statistical model of the probability distribution of the measurements. Thus, it is itself a random variable which can be constructed from a probability distribution as a function of transverse deflections. For any such estimator, classical estimation theory states that the standard deviation is lower bounded by \cite{helstrom1969quantum,FUJIWARA1995119}
\begin{equation}
	\text{Var}[\widetilde{\Delta \theta}]\geqslant\frac{1}{NF(\Delta \theta)},
\end{equation}
where $F(\Delta \theta)$ is the Fisher information, which quantifies the information that a specific measurement can reveal about an unknown target parameter, and can be calculated as
\begin{equation}
	\begin{aligned}
	F(\Delta \theta)=&\int{d}\Delta k\frac{(\partial _{\Delta \theta}P_0(\Delta k,\Delta \theta))^2}{P_0(\Delta k,\Delta \theta)}+\frac{(\partial _{\Delta \theta}P_1(\Delta k,\Delta \theta))^2}{P_1(\Delta k,\Delta \theta)}\\
&+\frac{(\partial _{\Delta \theta}P_2(\Delta k,\Delta \theta))^2}{P_2(\Delta k,\Delta \theta)} 
	\end{aligned}
\end{equation}
This limit is known as Cram$\acute{e}$r-Rao bound, which is tied to not only the probe state and interaction process, but also the specific measurement strategy. Evaluating the Fisher information for this set of probabilities, we find that its upper bound is
achieved in ideal case ($\gamma=0, \nu=1$) as
\begin{equation}
	F(\Delta \theta)\leqslant\int d\Delta k d^2 \Delta k^2 C(\Delta k)=2\sigma_k^2 d^2.
\end{equation}
In the case of zero loss and perfect visibility, the scheme recovers the quantum Cramér-Rao bound, thus confirming that the measurement strategy is indeed optimal. Furthermore, these statistical predictions suggest that the optimal sensing position in the presence of imperfect visibility or photon loss depends on specific experimental parameters \cite{chen2019hong,lyons2018attoseond}. Thus, our scheme also provides a powerful tool to determine the optimum working points for ultimate sensing precision.

\section{Experimental implementation}
\begin{figure*}
	\centering
	\includegraphics[width=1\linewidth]{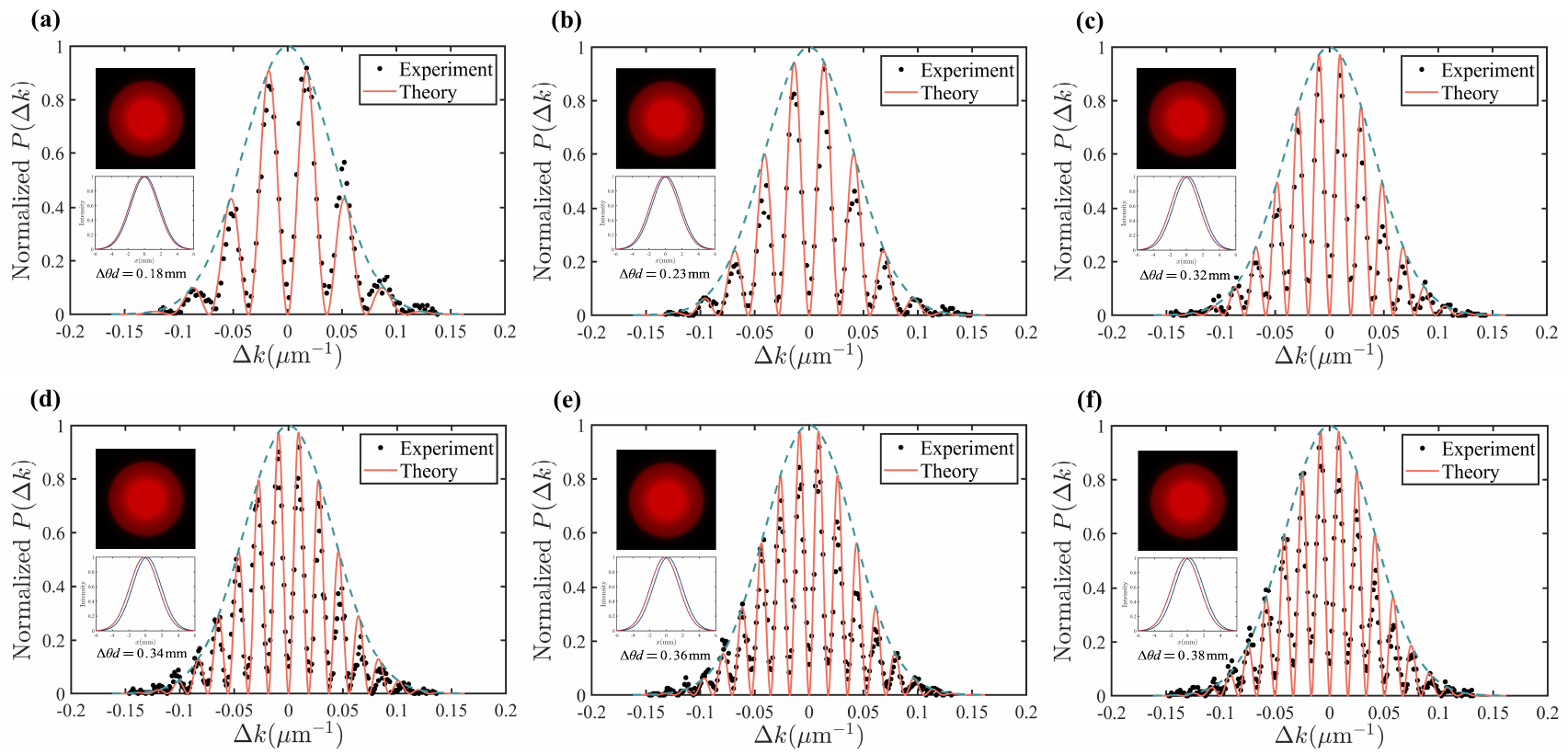}
	\caption{ Experimental measurements of the normalized joint probabilities for (a) $\Delta \theta=0.52$ \text{mrad}, (b) $\Delta \theta=0.67$ \text{mrad}, (c) $\Delta \theta=0.96$ \text{mrad}, (d) $\Delta \theta=1.01$ \text{mrad}, (e) $\Delta \theta=1.06$ \text{mrad}, and (f) $\Delta \theta=1.12$ \text{mrad}, where the orange
lines represent the theoretical predictions, and the black points represent the experimental results. The insets shown in the left side are their corresponding schematics of the spatial distributions from the paired photons in the detection screens. For the sake of clarity, their normalized intensities are also plotted to demonstrate the tiny transverse deflection.}
    \label{fig3}
\end{figure*}
We demonstrate a spatially resolved HOM interferometry that is used to detect transverse deflection introduced by a slight tilt of one photonic beam as shown in Fig. \ref{fig2}. The photon pairs are generated via spontaneous parametric down-conversion (SPDC) process pumped with a continuous wave laser at center wavelength of 405 nm. A 5-mm long nonlinear PPKTP crystal with a grating period of 10.025 \textmu m for type-II quasi-phase-matching is used such that the paired photons have orthogonal polarization. By placing this nonlinear crystal in a Sagnac interferometer, both of the clockwise and counterclockwise directions are pumped equally when setting the incident pump beam at anti-diagonal polarization. After combing the photon pairs from both direction on a polarization beam splitter, a anti-symmetric polarization entanglement state can be created as $\ket{\phi}=(\ket{H}\ket{V}-\ket{V}\ket{H})/\sqrt{2}$, where $H$ and $V$ represent the horizontal and vertical polarization, respectively \cite{kim2006phase,2018chenpolarization}. For erasing the temporal distinguishable information between different input paths of the HOM interferometer, a motorized Linear Stage is used for temporal compensation. Then these photons are coupled into the single mode fibers to filter their spatial modes to a Gaussian function, and are routed into a balanced beam splitter from opposite input ports. Since the incident photons are entangled in an anti-symmetric quantum state, they would be anti-bunched into distinct output ports of the beam splitter, and thus the coincidence detection events are identified by two detectors at opposite spatial modes. We note that the collimators used in spatial HOM interferometry are elaborate and customized, who can make the beam size of the output light reach 12 mm. This is significantly beneficial to implement the spatially resolved coincidence detection. In order to scan the specific transverse momentum, a movable slit was placed in front of the beam collimators with a width of 150 \textmu m. Finally, the down-converted photons are detected by silicon avalanche photon diodes and twofold events are identified using a fast electronic AND gate when two photons arrive at the detectors within a coincidence window of approximately 1 ns. 

In our experimental measurement, the transverse deflections of $\Delta\theta=0.52$ \text{mrad}, $0.67$ \text{mrad}, $0.96$ \text{mrad}, $1.01$ \text{mrad}, $1.06$ \text{mrad}, and $1.12$ \text{mrad} are introduced by slightly rotating the collimator at the input port of HOM interferometry, and this precise rotation can be implemented by using a motorized rotation stage. However, as shown in the insets of Fig. \ref{fig3}, the spatial distributions of paired photons almost overlap completely, which cannot provide any useful information for extracting the transverse deflection by direct imaging technique. On the other hand, their resultant interference patterns are shown in Fig. \ref{fig3}. They clearly reveal the spatial oscillations within the Gaussian envelopes, whose oscillation period is shorter as the transverse deflection becomes larger. By fitting these interference patterns to normalized coincidence probability as shown in Eq. \eqref{eq: probability}, we are able to estimate the transverse-momentum distribution of single photons to be 0.029 $\mu$m$^{-1}$, which corresponds to a beam size of 12 mm. The achievable interference visibility in our experiment can reach 0.85$\pm$0.04. These experimental measurement results agree well with our theoretical prediction, where the slight deviation can be attributed to imperfect experimental components.

\begin{figure}
	\centering
	\includegraphics[width=1\linewidth]{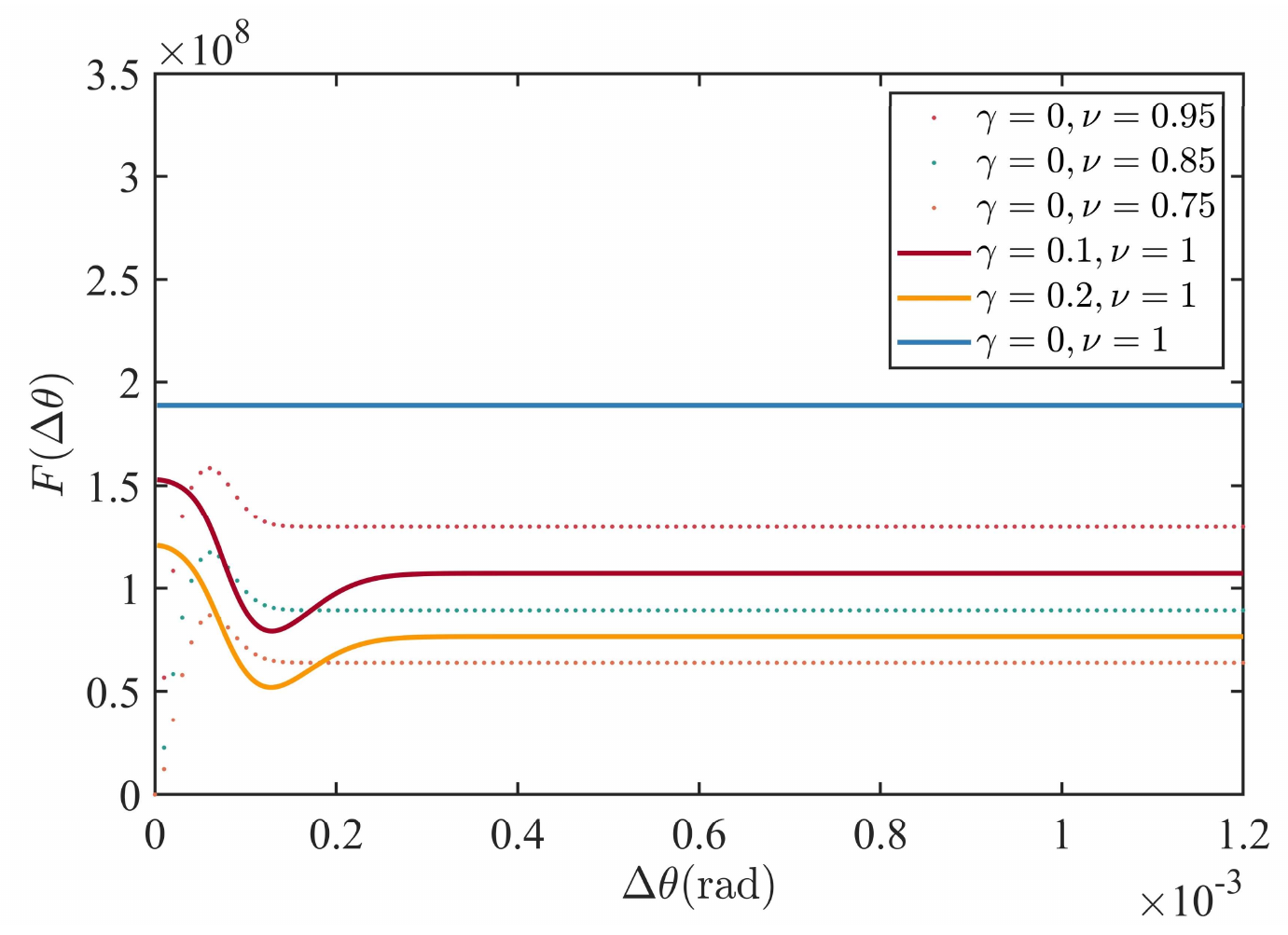}
	\caption{Theoretical prediction of the Fisher information as a function of the transverse deflection $\Delta \theta$ for various imperfect visibilities and channel-loss rates, wherein the standard deviation of the spatial distribution is $\sigma_k=0.029$ $\mu$m$^{-1}$, and the distance from the source to the detection is $d=335$ mm according to our experimental settings.}
    \label{fig4}
\end{figure}
In the context of our experimental condition, the ultimate Fisher information that can be achieved is shown in Fig. \ref{fig4}. No matter whether the experiment is subject to photon loss and imperfect visibility or not, the Fisher information is proportional to the spatial distribution of single photons and the distance between the source and the detector. In particular, when the experiment is subject to photon loss and imperfect visibility, the Fisher information is further relevant to the transverse deflection as shown in Fig. \ref{fig4}. Therefore, the values of parameters $\sigma_k$, $d$, $\Delta k$, $\nu$ and $\gamma$ need to be separately estimated before the measurements begin, and thus the optimum working pints can be determined by using a Fisher information analysis.

\section{DISCUSSION}
We have demonstrated an experimental approach to HOM interferometry for precise measurement of a transverse deflection, which is based on the two-photon interference and spatially resolved coincidence detection. In analogy to HOM interferometry on a biphoton beat note, the spatial beat pattern can pave the way to new high-precision sensing techniques. For example, in the condition of the spatial distribution of $\sigma_k$=0.029 $\mu$m$^{-1}$ and distance $d=335$ mm that have already realized in our experiment, a deflection sensitivity of 0.5 $\mu$rad can be obtained for only $N=10^4$ detection events. Backed by the results that our proof-of-principle experiment, this shows that the approach can provide higher resolution and highly sensitive measurement, makes it an ideal candidate for more quantum enhanced metrology applications. This work can be directly applied in high-precision refractometry and astrophysical bodies localization, where a tiny transverse deflection may be introduced by the unknown sample \cite{pang2008quantum,serbyn2014interferometric,barer1957refractometry}. Additionally, the spatial resolved HOM interferometry has the potential to enhance quantum imaging techniques, in particular for those require single-photon sources as the probe like the localization and tracking of biological samples \cite{ober2004localization,lelek2021single,hanne2015sted}. By combing this work with HOM interferometry in the frequency-time domain, it allows the simultaneous measurement of spatial and temporal unknown parameters with high precision, which has various applications in quantum microscopy \cite{ndagano2022quantum}, 3D quantum localization method \cite{persson2003quantum,Wang2021quantum}, and estimation of the color and position of single-photon emitters \cite{aharonovich2016solid}. We believe that fully harnessing spatially resolved HOM interference would provide additional tools to ultimately broaden the path towards practical quantum metrology.

\begin{acknowledgments}
This work is supported by the National Natural Science Foundation of China (12034016, 12205107), the National Key R$\&$D Program of China (2023YFA1407200), the Natural Science Foundation of Fujian Province of China (2021J02002), and the program for New Century Excellent Talents in University of China (NCET-13-0495).
\end{acknowledgments}
\bibliography{apssamp}

\providecommand{\noopsort}[1]{}\providecommand{\singleletter}[1]{#1}%
\begin{thebibliography}{48}%
\makeatletter
\providecommand \@ifxundefined [1]{%
 \@ifx{#1\undefined}
}%
\providecommand \@ifnum [1]{%
 \ifnum #1\expandafter \@firstoftwo
 \else \expandafter \@secondoftwo
 \fi
}%
\providecommand \@ifx [1]{%
 \ifx #1\expandafter \@firstoftwo
 \else \expandafter \@secondoftwo
 \fi
}%
\providecommand \natexlab [1]{#1}%
\providecommand \enquote  [1]{``#1''}%
\providecommand \bibnamefont  [1]{#1}%
\providecommand \bibfnamefont [1]{#1}%
\providecommand \citenamefont [1]{#1}%
\providecommand \href@noop [0]{\@secondoftwo}%
\providecommand \href [0]{\begingroup \@sanitize@url \@href}%
\providecommand \@href[1]{\@@startlink{#1}\@@href}%
\providecommand \@@href[1]{\endgroup#1\@@endlink}%
\providecommand \@sanitize@url [0]{\catcode `\\12\catcode `\$12\catcode
  `\&12\catcode `\#12\catcode `\^12\catcode `\_12\catcode `\%12\relax}%
\providecommand \@@startlink[1]{}%
\providecommand \@@endlink[0]{}%
\providecommand \url  [0]{\begingroup\@sanitize@url \@url }%
\providecommand \@url [1]{\endgroup\@href {#1}{\urlprefix }}%
\providecommand \urlprefix  [0]{URL }%
\providecommand \Eprint [0]{\href }%
\providecommand \doibase [0]{https://doi.org/}%
\providecommand \selectlanguage [0]{\@gobble}%
\providecommand \bibinfo  [0]{\@secondoftwo}%
\providecommand \bibfield  [0]{\@secondoftwo}%
\providecommand \translation [1]{[#1]}%
\providecommand \BibitemOpen [0]{}%
\providecommand \bibitemStop [0]{}%
\providecommand \bibitemNoStop [0]{.\EOS\space}%
\providecommand \EOS [0]{\spacefactor3000\relax}%
\providecommand \BibitemShut  [1]{\csname bibitem#1\endcsname}%
\let\auto@bib@innerbib\@empty
\bibitem [{\citenamefont {Giovannetti}\ \emph {et~al.}(2011)\citenamefont
  {Giovannetti}, \citenamefont {Lloyd},\ and\ \citenamefont
  {Maccone}}]{giovannetti2011advances}%
  \BibitemOpen
  \bibfield  {author} {\bibinfo {author} {\bibfnamefont {V.}~\bibnamefont
  {Giovannetti}}, \bibinfo {author} {\bibfnamefont {S.}~\bibnamefont {Lloyd}},\
  and\ \bibinfo {author} {\bibfnamefont {L.}~\bibnamefont {Maccone}},\
  }\bibfield  {title} {\bibinfo {title} {Advances in quantum metrology},\
  }\href {https://doi.org/10.1038/nphoton.2011.35} {\bibfield  {journal}
  {\bibinfo  {journal} {Nature photonics}\ }\textbf {\bibinfo {volume} {5}},\
  \bibinfo {pages} {222} (\bibinfo {year} {2011})}\BibitemShut {NoStop}%
\bibitem [{\citenamefont {Slussarenko}\ \emph {et~al.}(2017)\citenamefont
  {Slussarenko}, \citenamefont {Weston}, \citenamefont {Chrzanowski},
  \citenamefont {Shalm}, \citenamefont {Verma}, \citenamefont {Nam},\ and\
  \citenamefont {Pryde}}]{slussarenko2017unconditional}%
  \BibitemOpen
  \bibfield  {author} {\bibinfo {author} {\bibfnamefont {S.}~\bibnamefont
  {Slussarenko}}, \bibinfo {author} {\bibfnamefont {M.~M.}\ \bibnamefont
  {Weston}}, \bibinfo {author} {\bibfnamefont {H.~M.}\ \bibnamefont
  {Chrzanowski}}, \bibinfo {author} {\bibfnamefont {L.~K.}\ \bibnamefont
  {Shalm}}, \bibinfo {author} {\bibfnamefont {V.~B.}\ \bibnamefont {Verma}},
  \bibinfo {author} {\bibfnamefont {S.~W.}\ \bibnamefont {Nam}},\ and\ \bibinfo
  {author} {\bibfnamefont {G.~J.}\ \bibnamefont {Pryde}},\ }\bibfield  {title}
  {\bibinfo {title} {Unconditional violation of the shot-noise limit in
  photonic quantum metrology},\ }\href@noop {} {\bibfield  {journal} {\bibinfo
  {journal} {Nature Photonics}\ }\textbf {\bibinfo {volume} {11}},\ \bibinfo
  {pages} {700} (\bibinfo {year} {2017})}\BibitemShut {NoStop}%
\bibitem [{\citenamefont {Giovannetti}\ \emph {et~al.}(2004)\citenamefont
  {Giovannetti}, \citenamefont {Lloyd},\ and\ \citenamefont
  {Maccone}}]{giovannetti2004quantum}%
  \BibitemOpen
  \bibfield  {author} {\bibinfo {author} {\bibfnamefont {V.}~\bibnamefont
  {Giovannetti}}, \bibinfo {author} {\bibfnamefont {S.}~\bibnamefont {Lloyd}},\
  and\ \bibinfo {author} {\bibfnamefont {L.}~\bibnamefont {Maccone}},\
  }\bibfield  {title} {\bibinfo {title} {Quantum-enhanced measurements: Beating
  the standard quantum limit},\ }\href
  {https://doi.org/10.1126/science.1104149} {\bibfield  {journal} {\bibinfo
  {journal} {Science}\ }\textbf {\bibinfo {volume} {306}},\ \bibinfo {pages}
  {1330} (\bibinfo {year} {2004})}\BibitemShut {NoStop}%
\bibitem [{\citenamefont {Wolfgramm}\ \emph {et~al.}(2013)\citenamefont
  {Wolfgramm}, \citenamefont {Vitelli}, \citenamefont {Beduini}, \citenamefont
  {Godbout},\ and\ \citenamefont {Mitchell}}]{wolfgramm2013entanglement}%
  \BibitemOpen
  \bibfield  {author} {\bibinfo {author} {\bibfnamefont {F.}~\bibnamefont
  {Wolfgramm}}, \bibinfo {author} {\bibfnamefont {C.}~\bibnamefont {Vitelli}},
  \bibinfo {author} {\bibfnamefont {F.~A.}\ \bibnamefont {Beduini}}, \bibinfo
  {author} {\bibfnamefont {N.}~\bibnamefont {Godbout}},\ and\ \bibinfo {author}
  {\bibfnamefont {M.~W.}\ \bibnamefont {Mitchell}},\ }\bibfield  {title}
  {\bibinfo {title} {Entanglement-enhanced probing of a delicate material
  system},\ }\href {https://doi.org/10.1038/nphoton.2012.300} {\bibfield
  {journal} {\bibinfo  {journal} {Nature Photonics}\ }\textbf {\bibinfo
  {volume} {7}},\ \bibinfo {pages} {28} (\bibinfo {year} {2013})}\BibitemShut
  {NoStop}%
\bibitem [{\citenamefont {Holland}\ and\ \citenamefont
  {Burnett}(1993)}]{holland1993interferometric}%
  \BibitemOpen
  \bibfield  {author} {\bibinfo {author} {\bibfnamefont {M.~J.}\ \bibnamefont
  {Holland}}\ and\ \bibinfo {author} {\bibfnamefont {K.}~\bibnamefont
  {Burnett}},\ }\bibfield  {title} {\bibinfo {title} {Interferometric detection
  of optical phase shifts at the heisenberg limit},\ }\href
  {https://doi.org/10.1103/PhysRevLett.71.1355} {\bibfield  {journal} {\bibinfo
   {journal} {Phys. Rev. Lett.}\ }\textbf {\bibinfo {volume} {71}},\ \bibinfo
  {pages} {1355} (\bibinfo {year} {1993})}\BibitemShut {NoStop}%
\bibitem [{\citenamefont {Resch}\ \emph {et~al.}(2007)\citenamefont {Resch},
  \citenamefont {Pregnell}, \citenamefont {Prevedel}, \citenamefont
  {Gilchrist}, \citenamefont {Pryde}, \citenamefont {O'Brien},\ and\
  \citenamefont {White}}]{resch2007time}%
  \BibitemOpen
  \bibfield  {author} {\bibinfo {author} {\bibfnamefont {K.~J.}\ \bibnamefont
  {Resch}}, \bibinfo {author} {\bibfnamefont {K.~L.}\ \bibnamefont {Pregnell}},
  \bibinfo {author} {\bibfnamefont {R.}~\bibnamefont {Prevedel}}, \bibinfo
  {author} {\bibfnamefont {A.}~\bibnamefont {Gilchrist}}, \bibinfo {author}
  {\bibfnamefont {G.~J.}\ \bibnamefont {Pryde}}, \bibinfo {author}
  {\bibfnamefont {J.~L.}\ \bibnamefont {O'Brien}},\ and\ \bibinfo {author}
  {\bibfnamefont {A.~G.}\ \bibnamefont {White}},\ }\bibfield  {title} {\bibinfo
  {title} {Time-reversal and super-resolving phase measurements},\ }\href
  {https://doi.org/10.1103/PhysRevLett.98.223601} {\bibfield  {journal}
  {\bibinfo  {journal} {Phys. Rev. Lett.}\ }\textbf {\bibinfo {volume} {98}},\
  \bibinfo {pages} {223601} (\bibinfo {year} {2007})}\BibitemShut {NoStop}%
\bibitem [{\citenamefont {Mitchell}\ \emph {et~al.}(2004)\citenamefont
  {Mitchell}, \citenamefont {Lundeen},\ and\ \citenamefont
  {Steinberg}}]{mitchell2004super}%
  \BibitemOpen
  \bibfield  {author} {\bibinfo {author} {\bibfnamefont {M.~W.}\ \bibnamefont
  {Mitchell}}, \bibinfo {author} {\bibfnamefont {J.~S.}\ \bibnamefont
  {Lundeen}},\ and\ \bibinfo {author} {\bibfnamefont {A.~M.}\ \bibnamefont
  {Steinberg}},\ }\bibfield  {title} {\bibinfo {title} {Super-resolving phase
  measurements with a multiphoton entangled state},\ }\href
  {https://doi.org/10.1038/nature02493} {\bibfield  {journal} {\bibinfo
  {journal} {Nature}\ }\textbf {\bibinfo {volume} {429}},\ \bibinfo {pages}
  {161} (\bibinfo {year} {2004})}\BibitemShut {NoStop}%
\bibitem [{\citenamefont {Nagata}\ \emph {et~al.}(2007)\citenamefont {Nagata},
  \citenamefont {Okamoto}, \citenamefont {O'Brien}, \citenamefont {Sasaki},\
  and\ \citenamefont {Takeuchi}}]{nagata2007beating}%
  \BibitemOpen
  \bibfield  {author} {\bibinfo {author} {\bibfnamefont {T.}~\bibnamefont
  {Nagata}}, \bibinfo {author} {\bibfnamefont {R.}~\bibnamefont {Okamoto}},
  \bibinfo {author} {\bibfnamefont {J.~L.}\ \bibnamefont {O'Brien}}, \bibinfo
  {author} {\bibfnamefont {K.}~\bibnamefont {Sasaki}},\ and\ \bibinfo {author}
  {\bibfnamefont {S.}~\bibnamefont {Takeuchi}},\ }\bibfield  {title} {\bibinfo
  {title} {Beating the standard quantum limit with four-entangled photons},\
  }\href {https://doi.org/10.1126/science.1138007} {\bibfield  {journal}
  {\bibinfo  {journal} {Science}\ }\textbf {\bibinfo {volume} {316}},\ \bibinfo
  {pages} {726} (\bibinfo {year} {2007})}\BibitemShut {NoStop}%
\bibitem [{\citenamefont {Hong}\ \emph {et~al.}(1987)\citenamefont {Hong},
  \citenamefont {Ou},\ and\ \citenamefont {Mandel}}]{hong1987measurement}%
  \BibitemOpen
  \bibfield  {author} {\bibinfo {author} {\bibfnamefont {C.~K.}\ \bibnamefont
  {Hong}}, \bibinfo {author} {\bibfnamefont {Z.~Y.}\ \bibnamefont {Ou}},\ and\
  \bibinfo {author} {\bibfnamefont {L.}~\bibnamefont {Mandel}},\ }\bibfield
  {title} {\bibinfo {title} {Measurement of subpicosecond time intervals
  between two photons by interference},\ }\href
  {https://doi.org/10.1103/PhysRevLett.59.2044} {\bibfield  {journal} {\bibinfo
   {journal} {Phys. Rev. Lett.}\ }\textbf {\bibinfo {volume} {59}},\ \bibinfo
  {pages} {2044} (\bibinfo {year} {1987})}\BibitemShut {NoStop}%
\bibitem [{\citenamefont {Lyons}\ \emph {et~al.}(2018)\citenamefont {Lyons},
  \citenamefont {Knee}, \citenamefont {Bolduc}, \citenamefont {Roger},
  \citenamefont {Leach}, \citenamefont {Gauger},\ and\ \citenamefont
  {Faccio}}]{lyons2018attoseond}%
  \BibitemOpen
  \bibfield  {author} {\bibinfo {author} {\bibfnamefont {A.}~\bibnamefont
  {Lyons}}, \bibinfo {author} {\bibfnamefont {G.~C.}\ \bibnamefont {Knee}},
  \bibinfo {author} {\bibfnamefont {E.}~\bibnamefont {Bolduc}}, \bibinfo
  {author} {\bibfnamefont {T.}~\bibnamefont {Roger}}, \bibinfo {author}
  {\bibfnamefont {J.}~\bibnamefont {Leach}}, \bibinfo {author} {\bibfnamefont
  {E.~M.}\ \bibnamefont {Gauger}},\ and\ \bibinfo {author} {\bibfnamefont
  {D.}~\bibnamefont {Faccio}},\ }\bibfield  {title} {\bibinfo {title}
  {Attosecond-resolution hong-ou-mandel interferometry},\ }\href
  {https://doi.org/10.1126/sciadv.aap9416} {\bibfield  {journal} {\bibinfo
  {journal} {Science Advances}\ }\textbf {\bibinfo {volume} {4}},\ \bibinfo
  {pages} {eaap9416} (\bibinfo {year} {2018})}\BibitemShut {NoStop}%
\bibitem [{\citenamefont {Chen}\ \emph {et~al.}(2019)\citenamefont {Chen},
  \citenamefont {Fink}, \citenamefont {Steinlechner}, \citenamefont {Torres},\
  and\ \citenamefont {Ursin}}]{chen2019hong}%
  \BibitemOpen
  \bibfield  {author} {\bibinfo {author} {\bibfnamefont {Y.}~\bibnamefont
  {Chen}}, \bibinfo {author} {\bibfnamefont {M.}~\bibnamefont {Fink}}, \bibinfo
  {author} {\bibfnamefont {F.}~\bibnamefont {Steinlechner}}, \bibinfo {author}
  {\bibfnamefont {J.~P.}\ \bibnamefont {Torres}},\ and\ \bibinfo {author}
  {\bibfnamefont {R.}~\bibnamefont {Ursin}},\ }\bibfield  {title} {\bibinfo
  {title} {Hong-ou-mandel interferometry on a biphoton beat note},\ }\href
  {https://doi.org/10.1038/s41534-019-0161-z} {\bibfield  {journal} {\bibinfo
  {journal} {npj Quantum Information}\ }\textbf {\bibinfo {volume} {5}},\
  \bibinfo {pages} {43} (\bibinfo {year} {2019})}\BibitemShut {NoStop}%
\bibitem [{\citenamefont {Nasr}\ \emph {et~al.}(2003)\citenamefont {Nasr},
  \citenamefont {Saleh}, \citenamefont {Sergienko},\ and\ \citenamefont
  {Teich}}]{nasr2003demonstration}%
  \BibitemOpen
  \bibfield  {author} {\bibinfo {author} {\bibfnamefont {M.~B.}\ \bibnamefont
  {Nasr}}, \bibinfo {author} {\bibfnamefont {B.~E.~A.}\ \bibnamefont {Saleh}},
  \bibinfo {author} {\bibfnamefont {A.~V.}\ \bibnamefont {Sergienko}},\ and\
  \bibinfo {author} {\bibfnamefont {M.~C.}\ \bibnamefont {Teich}},\ }\bibfield
  {title} {\bibinfo {title} {Demonstration of dispersion-canceled
  quantum-optical coherence tomography},\ }\href
  {https://doi.org/10.1103/PhysRevLett.91.083601} {\bibfield  {journal}
  {\bibinfo  {journal} {Phys. Rev. Lett.}\ }\textbf {\bibinfo {volume} {91}},\
  \bibinfo {pages} {083601} (\bibinfo {year} {2003})}\BibitemShut {NoStop}%
\bibitem [{\citenamefont {Ibarra-Borja}\ \emph {et~al.}(2020)\citenamefont
  {Ibarra-Borja}, \citenamefont {Sevilla-Guti\'{e}rrez}, \citenamefont
  {Ram\'{i}rez-Alarc\'{o}n}, \citenamefont {Cruz-Ram\'{i}rez},\ and\
  \citenamefont {U'Ren}}]{Ibarra-Borja:20}%
  \BibitemOpen
  \bibfield  {author} {\bibinfo {author} {\bibfnamefont {Z.}~\bibnamefont
  {Ibarra-Borja}}, \bibinfo {author} {\bibfnamefont {C.}~\bibnamefont
  {Sevilla-Guti\'{e}rrez}}, \bibinfo {author} {\bibfnamefont {R.}~\bibnamefont
  {Ram\'{i}rez-Alarc\'{o}n}}, \bibinfo {author} {\bibfnamefont
  {H.}~\bibnamefont {Cruz-Ram\'{i}rez}},\ and\ \bibinfo {author} {\bibfnamefont
  {A.~B.}\ \bibnamefont {U'Ren}},\ }\bibfield  {title} {\bibinfo {title}
  {Experimental demonstration of full-field quantum optical coherence
  tomography},\ }\href {https://doi.org/10.1364/PRJ.8.000051} {\bibfield
  {journal} {\bibinfo  {journal} {Photon. Res.}\ }\textbf {\bibinfo {volume}
  {8}},\ \bibinfo {pages} {51} (\bibinfo {year} {2020})}\BibitemShut {NoStop}%
\bibitem [{\citenamefont {Chen}\ \emph {et~al.}(2023)\citenamefont {Chen},
  \citenamefont {Chen},\ and\ \citenamefont {Chen}}]{chen2023spectrally}%
  \BibitemOpen
  \bibfield  {author} {\bibinfo {author} {\bibfnamefont {C.}~\bibnamefont
  {Chen}}, \bibinfo {author} {\bibfnamefont {Y.}~\bibnamefont {Chen}},\ and\
  \bibinfo {author} {\bibfnamefont {L.}~\bibnamefont {Chen}},\ }\bibfield
  {title} {\bibinfo {title} {Spectrally resolved hong-ou-mandel interferometry
  with discrete color entanglement},\ }\href
  {https://doi.org/10.1103/PhysRevApplied.19.054092} {\bibfield  {journal}
  {\bibinfo  {journal} {Phys. Rev. Appl.}\ }\textbf {\bibinfo {volume} {19}},\
  \bibinfo {pages} {054092} (\bibinfo {year} {2023})}\BibitemShut {NoStop}%
\bibitem [{\citenamefont {Ndagano}\ \emph {et~al.}(2022)\citenamefont
  {Ndagano}, \citenamefont {Defienne}, \citenamefont {Branford}, \citenamefont
  {Shah}, \citenamefont {Lyons}, \citenamefont {Westerberg}, \citenamefont
  {Gauger},\ and\ \citenamefont {Faccio}}]{ndagano2022quantum}%
  \BibitemOpen
  \bibfield  {author} {\bibinfo {author} {\bibfnamefont {B.}~\bibnamefont
  {Ndagano}}, \bibinfo {author} {\bibfnamefont {H.}~\bibnamefont {Defienne}},
  \bibinfo {author} {\bibfnamefont {D.}~\bibnamefont {Branford}}, \bibinfo
  {author} {\bibfnamefont {Y.~D.}\ \bibnamefont {Shah}}, \bibinfo {author}
  {\bibfnamefont {A.}~\bibnamefont {Lyons}}, \bibinfo {author} {\bibfnamefont
  {N.}~\bibnamefont {Westerberg}}, \bibinfo {author} {\bibfnamefont {E.~M.}\
  \bibnamefont {Gauger}},\ and\ \bibinfo {author} {\bibfnamefont
  {D.}~\bibnamefont {Faccio}},\ }\bibfield  {title} {\bibinfo {title} {Quantum
  microscopy based on hong--ou--mandel interference},\ }\href
  {https://doi.org/10.1038/s41566-022-00980-6} {\bibfield  {journal} {\bibinfo
  {journal} {Nature Photonics}\ }\textbf {\bibinfo {volume} {16}},\ \bibinfo
  {pages} {384} (\bibinfo {year} {2022})}\BibitemShut {NoStop}%
\bibitem [{\citenamefont {Devaux}\ \emph {et~al.}(2020)\citenamefont {Devaux},
  \citenamefont {Mosset}, \citenamefont {Moreau},\ and\ \citenamefont
  {Lantz}}]{devaux2020imaging}%
  \BibitemOpen
  \bibfield  {author} {\bibinfo {author} {\bibfnamefont {F.}~\bibnamefont
  {Devaux}}, \bibinfo {author} {\bibfnamefont {A.}~\bibnamefont {Mosset}},
  \bibinfo {author} {\bibfnamefont {P.-A.}\ \bibnamefont {Moreau}},\ and\
  \bibinfo {author} {\bibfnamefont {E.}~\bibnamefont {Lantz}},\ }\bibfield
  {title} {\bibinfo {title} {Imaging spatiotemporal hong-ou-mandel interference
  of biphoton states of extremely high schmidt number},\ }\href
  {https://doi.org/10.1103/PhysRevX.10.031031} {\bibfield  {journal} {\bibinfo
  {journal} {Phys. Rev. X}\ }\textbf {\bibinfo {volume} {10}},\ \bibinfo
  {pages} {031031} (\bibinfo {year} {2020})}\BibitemShut {NoStop}%
\bibitem [{\citenamefont {Huang}\ \emph {et~al.}(1991)\citenamefont {Huang},
  \citenamefont {Swanson}, \citenamefont {Lin}, \citenamefont {Schuman},
  \citenamefont {Stinson}, \citenamefont {Chang}, \citenamefont {Hee},
  \citenamefont {Flotte}, \citenamefont {Gregory}, \citenamefont {Puliafito},\
  and\ \citenamefont {Fujimoto}}]{devid1991oprical}%
  \BibitemOpen
  \bibfield  {author} {\bibinfo {author} {\bibfnamefont {D.}~\bibnamefont
  {Huang}}, \bibinfo {author} {\bibfnamefont {E.~A.}\ \bibnamefont {Swanson}},
  \bibinfo {author} {\bibfnamefont {C.~P.}\ \bibnamefont {Lin}}, \bibinfo
  {author} {\bibfnamefont {J.~S.}\ \bibnamefont {Schuman}}, \bibinfo {author}
  {\bibfnamefont {W.~G.}\ \bibnamefont {Stinson}}, \bibinfo {author}
  {\bibfnamefont {W.}~\bibnamefont {Chang}}, \bibinfo {author} {\bibfnamefont
  {M.~R.}\ \bibnamefont {Hee}}, \bibinfo {author} {\bibfnamefont
  {T.}~\bibnamefont {Flotte}}, \bibinfo {author} {\bibfnamefont
  {K.}~\bibnamefont {Gregory}}, \bibinfo {author} {\bibfnamefont {C.~A.}\
  \bibnamefont {Puliafito}},\ and\ \bibinfo {author} {\bibfnamefont {J.~G.}\
  \bibnamefont {Fujimoto}},\ }\bibfield  {title} {\bibinfo {title} {Optical
  coherence tomography},\ }\href {https://doi.org/10.1126/science.1957169}
  {\bibfield  {journal} {\bibinfo  {journal} {Science}\ }\textbf {\bibinfo
  {volume} {254}},\ \bibinfo {pages} {1178} (\bibinfo {year}
  {1991})}\BibitemShut {NoStop}%
\bibitem [{\citenamefont {Bouma}\ \emph {et~al.}(2022)\citenamefont {Bouma},
  \citenamefont {de~Boer}, \citenamefont {Huang}, \citenamefont {Jang},
  \citenamefont {Yonetsu}, \citenamefont {Leggett}, \citenamefont {Leitgeb},
  \citenamefont {Sampson}, \citenamefont {Suter}, \citenamefont {Vakoc} \emph
  {et~al.}}]{bouma2022optical}%
  \BibitemOpen
  \bibfield  {author} {\bibinfo {author} {\bibfnamefont {B.~E.}\ \bibnamefont
  {Bouma}}, \bibinfo {author} {\bibfnamefont {J.~F.}\ \bibnamefont {de~Boer}},
  \bibinfo {author} {\bibfnamefont {D.}~\bibnamefont {Huang}}, \bibinfo
  {author} {\bibfnamefont {I.-K.}\ \bibnamefont {Jang}}, \bibinfo {author}
  {\bibfnamefont {T.}~\bibnamefont {Yonetsu}}, \bibinfo {author} {\bibfnamefont
  {C.~L.}\ \bibnamefont {Leggett}}, \bibinfo {author} {\bibfnamefont
  {R.}~\bibnamefont {Leitgeb}}, \bibinfo {author} {\bibfnamefont {D.~D.}\
  \bibnamefont {Sampson}}, \bibinfo {author} {\bibfnamefont {M.}~\bibnamefont
  {Suter}}, \bibinfo {author} {\bibfnamefont {B.~J.}\ \bibnamefont {Vakoc}},
  \emph {et~al.},\ }\bibfield  {title} {\bibinfo {title} {Optical coherence
  tomography},\ }\href {https://doi.org/10.1038/s43586-022-00162-2} {\bibfield
  {journal} {\bibinfo  {journal} {Nature Reviews Methods Primers}\ }\textbf
  {\bibinfo {volume} {2}},\ \bibinfo {pages} {79} (\bibinfo {year}
  {2022})}\BibitemShut {NoStop}%
\bibitem [{\citenamefont {Legero}\ \emph {et~al.}(2004)\citenamefont {Legero},
  \citenamefont {Wilk}, \citenamefont {Hennrich}, \citenamefont {Rempe},\ and\
  \citenamefont {Kuhn}}]{Legero2004quantum}%
  \BibitemOpen
  \bibfield  {author} {\bibinfo {author} {\bibfnamefont {T.}~\bibnamefont
  {Legero}}, \bibinfo {author} {\bibfnamefont {T.}~\bibnamefont {Wilk}},
  \bibinfo {author} {\bibfnamefont {M.}~\bibnamefont {Hennrich}}, \bibinfo
  {author} {\bibfnamefont {G.}~\bibnamefont {Rempe}},\ and\ \bibinfo {author}
  {\bibfnamefont {A.}~\bibnamefont {Kuhn}},\ }\bibfield  {title} {\bibinfo
  {title} {Quantum beat of two single photons},\ }\href
  {https://doi.org/10.1103/PhysRevLett.93.070503} {\bibfield  {journal}
  {\bibinfo  {journal} {Phys. Rev. Lett.}\ }\textbf {\bibinfo {volume} {93}},\
  \bibinfo {pages} {070503} (\bibinfo {year} {2004})}\BibitemShut {NoStop}%
\bibitem [{\citenamefont {Ramelow}\ \emph {et~al.}(2009)\citenamefont
  {Ramelow}, \citenamefont {Ratschbacher}, \citenamefont {Fedrizzi},
  \citenamefont {Langford},\ and\ \citenamefont
  {Zeilinger}}]{ramelow2009discrete}%
  \BibitemOpen
  \bibfield  {author} {\bibinfo {author} {\bibfnamefont {S.}~\bibnamefont
  {Ramelow}}, \bibinfo {author} {\bibfnamefont {L.}~\bibnamefont
  {Ratschbacher}}, \bibinfo {author} {\bibfnamefont {A.}~\bibnamefont
  {Fedrizzi}}, \bibinfo {author} {\bibfnamefont {N.~K.}\ \bibnamefont
  {Langford}},\ and\ \bibinfo {author} {\bibfnamefont {A.}~\bibnamefont
  {Zeilinger}},\ }\bibfield  {title} {\bibinfo {title} {Discrete tunable color
  entanglement},\ }\href {https://doi.org/10.1103/PhysRevLett.103.253601}
  {\bibfield  {journal} {\bibinfo  {journal} {Phys. Rev. Lett.}\ }\textbf
  {\bibinfo {volume} {103}},\ \bibinfo {pages} {253601} (\bibinfo {year}
  {2009})}\BibitemShut {NoStop}%
\bibitem [{\citenamefont {Tamma}\ and\ \citenamefont
  {Laibacher}(2015)}]{tamma2015multiboson}%
  \BibitemOpen
  \bibfield  {author} {\bibinfo {author} {\bibfnamefont {V.}~\bibnamefont
  {Tamma}}\ and\ \bibinfo {author} {\bibfnamefont {S.}~\bibnamefont
  {Laibacher}},\ }\bibfield  {title} {\bibinfo {title} {Multiboson correlation
  interferometry with arbitrary single-photon pure states},\ }\href
  {https://doi.org/10.1103/PhysRevLett.114.243601} {\bibfield  {journal}
  {\bibinfo  {journal} {Phys. Rev. Lett.}\ }\textbf {\bibinfo {volume} {114}},\
  \bibinfo {pages} {243601} (\bibinfo {year} {2015})}\BibitemShut {NoStop}%
\bibitem [{\citenamefont {Marchand}\ \emph {et~al.}(2021)\citenamefont
  {Marchand}, \citenamefont {Riemensberger}, \citenamefont {Skehan},
  \citenamefont {Ho}, \citenamefont {Pfeiffer}, \citenamefont {Liu},
  \citenamefont {Hauger}, \citenamefont {Lasser},\ and\ \citenamefont
  {Kippenberg}}]{marchand2021soliton}%
  \BibitemOpen
  \bibfield  {author} {\bibinfo {author} {\bibfnamefont {P.~J.}\ \bibnamefont
  {Marchand}}, \bibinfo {author} {\bibfnamefont {J.}~\bibnamefont
  {Riemensberger}}, \bibinfo {author} {\bibfnamefont {J.~C.}\ \bibnamefont
  {Skehan}}, \bibinfo {author} {\bibfnamefont {J.-J.}\ \bibnamefont {Ho}},
  \bibinfo {author} {\bibfnamefont {M.~H.}\ \bibnamefont {Pfeiffer}}, \bibinfo
  {author} {\bibfnamefont {J.}~\bibnamefont {Liu}}, \bibinfo {author}
  {\bibfnamefont {C.}~\bibnamefont {Hauger}}, \bibinfo {author} {\bibfnamefont
  {T.}~\bibnamefont {Lasser}},\ and\ \bibinfo {author} {\bibfnamefont {T.~J.}\
  \bibnamefont {Kippenberg}},\ }\bibfield  {title} {\bibinfo {title} {Soliton
  microcomb based spectral domain optical coherence tomography},\ }\href
  {https://doi.org/10.1038/s41467-020-20404-9} {\bibfield  {journal} {\bibinfo
  {journal} {Nature Communications}\ }\textbf {\bibinfo {volume} {12}},\
  \bibinfo {pages} {427} (\bibinfo {year} {2021})}\BibitemShut {NoStop}%
\bibitem [{\citenamefont {Leitgeb}\ \emph {et~al.}(2003)\citenamefont
  {Leitgeb}, \citenamefont {Hitzenberger},\ and\ \citenamefont
  {Fercher}}]{Leitgeb2003optical}%
  \BibitemOpen
  \bibfield  {author} {\bibinfo {author} {\bibfnamefont {R.}~\bibnamefont
  {Leitgeb}}, \bibinfo {author} {\bibfnamefont {C.~K.}\ \bibnamefont
  {Hitzenberger}},\ and\ \bibinfo {author} {\bibfnamefont {A.~F.}\ \bibnamefont
  {Fercher}},\ }\bibfield  {title} {\bibinfo {title} {Performance of fourier
  domain vs. time domain optical coherence tomography},\ }\href
  {https://doi.org/10.1364/OE.11.000889} {\bibfield  {journal} {\bibinfo
  {journal} {Opt. Express}\ }\textbf {\bibinfo {volume} {11}},\ \bibinfo
  {pages} {889} (\bibinfo {year} {2003})}\BibitemShut {NoStop}%
\bibitem [{\citenamefont {Rank}\ \emph {et~al.}(2021)\citenamefont {Rank},
  \citenamefont {Sentosa}, \citenamefont {Harper}, \citenamefont {Salas},
  \citenamefont {Gaugutz}, \citenamefont {Seyringer}, \citenamefont
  {Nevlacsil}, \citenamefont {Maese-Novo}, \citenamefont {Eggeling},
  \citenamefont {Muellner} \emph {et~al.}}]{rank2021toward}%
  \BibitemOpen
  \bibfield  {author} {\bibinfo {author} {\bibfnamefont {E.~A.}\ \bibnamefont
  {Rank}}, \bibinfo {author} {\bibfnamefont {R.}~\bibnamefont {Sentosa}},
  \bibinfo {author} {\bibfnamefont {D.~J.}\ \bibnamefont {Harper}}, \bibinfo
  {author} {\bibfnamefont {M.}~\bibnamefont {Salas}}, \bibinfo {author}
  {\bibfnamefont {A.}~\bibnamefont {Gaugutz}}, \bibinfo {author} {\bibfnamefont
  {D.}~\bibnamefont {Seyringer}}, \bibinfo {author} {\bibfnamefont
  {S.}~\bibnamefont {Nevlacsil}}, \bibinfo {author} {\bibfnamefont
  {A.}~\bibnamefont {Maese-Novo}}, \bibinfo {author} {\bibfnamefont
  {M.}~\bibnamefont {Eggeling}}, \bibinfo {author} {\bibfnamefont
  {P.}~\bibnamefont {Muellner}}, \emph {et~al.},\ }\bibfield  {title} {\bibinfo
  {title} {Toward optical coherence tomography on a chip: in vivo
  three-dimensional human retinal imaging using photonic integrated
  circuit-based arrayed waveguide gratings},\ }\href
  {https://doi.org/10.1038/s41377-020-00450-0} {\bibfield  {journal} {\bibinfo
  {journal} {Light: Science \& Applications}\ }\textbf {\bibinfo {volume}
  {10}},\ \bibinfo {pages} {6} (\bibinfo {year} {2021})}\BibitemShut {NoStop}%
\bibitem [{\citenamefont {Song}\ \emph {et~al.}(2021)\citenamefont {Song},
  \citenamefont {Jelly}, \citenamefont {Chu}, \citenamefont {Kendall},\ and\
  \citenamefont {Wax}}]{Song2021review}%
  \BibitemOpen
  \bibfield  {author} {\bibinfo {author} {\bibfnamefont {G.}~\bibnamefont
  {Song}}, \bibinfo {author} {\bibfnamefont {E.~T.}\ \bibnamefont {Jelly}},
  \bibinfo {author} {\bibfnamefont {K.~K.}\ \bibnamefont {Chu}}, \bibinfo
  {author} {\bibfnamefont {W.~Y.}\ \bibnamefont {Kendall}},\ and\ \bibinfo
  {author} {\bibfnamefont {A.}~\bibnamefont {Wax}},\ }\bibfield  {title}
  {\bibinfo {title} {A review of low-cost and portable optical coherence
  tomography},\ }\href {https://doi.org/10.1088/2516-1091/abfeb7} {\bibfield
  {journal} {\bibinfo  {journal} {Progress in Biomedical Engineering}\ }\textbf
  {\bibinfo {volume} {3}},\ \bibinfo {pages} {032002} (\bibinfo {year}
  {2021})}\BibitemShut {NoStop}%
\bibitem [{\citenamefont {Ou}\ and\ \citenamefont
  {Mandel}(1989)}]{ou1989further}%
  \BibitemOpen
  \bibfield  {author} {\bibinfo {author} {\bibfnamefont {Z.~Y.}\ \bibnamefont
  {Ou}}\ and\ \bibinfo {author} {\bibfnamefont {L.}~\bibnamefont {Mandel}},\
  }\bibfield  {title} {\bibinfo {title} {Further evidence of nonclassical
  behavior in optical interference},\ }\href
  {https://doi.org/10.1103/PhysRevLett.62.2941} {\bibfield  {journal} {\bibinfo
   {journal} {Phys. Rev. Lett.}\ }\textbf {\bibinfo {volume} {62}},\ \bibinfo
  {pages} {2941} (\bibinfo {year} {1989})}\BibitemShut {NoStop}%
\bibitem [{\citenamefont {Kim}\ \emph {et~al.}(2006{\natexlab{a}})\citenamefont
  {Kim}, \citenamefont {Kwon}, \citenamefont {Kim},\ and\ \citenamefont
  {Kim}}]{kim2006spatial}%
  \BibitemOpen
  \bibfield  {author} {\bibinfo {author} {\bibfnamefont {H.}~\bibnamefont
  {Kim}}, \bibinfo {author} {\bibfnamefont {O.}~\bibnamefont {Kwon}}, \bibinfo
  {author} {\bibfnamefont {W.}~\bibnamefont {Kim}},\ and\ \bibinfo {author}
  {\bibfnamefont {T.}~\bibnamefont {Kim}},\ }\bibfield  {title} {\bibinfo
  {title} {Spatial two-photon interference in a hong-ou-mandel
  interferometer},\ }\href {https://doi.org/10.1103/PhysRevA.73.023820}
  {\bibfield  {journal} {\bibinfo  {journal} {Phys. Rev. A}\ }\textbf {\bibinfo
  {volume} {73}},\ \bibinfo {pages} {023820} (\bibinfo {year}
  {2006}{\natexlab{a}})}\BibitemShut {NoStop}%
\bibitem [{\citenamefont {Lee}\ and\ \citenamefont {van
  Exter}(2006)}]{lee2006spatial}%
  \BibitemOpen
  \bibfield  {author} {\bibinfo {author} {\bibfnamefont {P.~S.~K.}\
  \bibnamefont {Lee}}\ and\ \bibinfo {author} {\bibfnamefont {M.~P.}\
  \bibnamefont {van Exter}},\ }\bibfield  {title} {\bibinfo {title} {Spatial
  labeling in a two-photon interferometer},\ }\href
  {https://doi.org/10.1103/PhysRevA.73.063827} {\bibfield  {journal} {\bibinfo
  {journal} {Phys. Rev. A}\ }\textbf {\bibinfo {volume} {73}},\ \bibinfo
  {pages} {063827} (\bibinfo {year} {2006})}\BibitemShut {NoStop}%
\bibitem [{\citenamefont {Hanne}\ \emph {et~al.}(2015)\citenamefont {Hanne},
  \citenamefont {Falk}, \citenamefont {G{\"o}rlitz}, \citenamefont {Hoyer},
  \citenamefont {Engelhardt}, \citenamefont {Sahl},\ and\ \citenamefont
  {Hell}}]{hanne2015sted}%
  \BibitemOpen
  \bibfield  {author} {\bibinfo {author} {\bibfnamefont {J.}~\bibnamefont
  {Hanne}}, \bibinfo {author} {\bibfnamefont {H.~J.}\ \bibnamefont {Falk}},
  \bibinfo {author} {\bibfnamefont {F.}~\bibnamefont {G{\"o}rlitz}}, \bibinfo
  {author} {\bibfnamefont {P.}~\bibnamefont {Hoyer}}, \bibinfo {author}
  {\bibfnamefont {J.}~\bibnamefont {Engelhardt}}, \bibinfo {author}
  {\bibfnamefont {S.~J.}\ \bibnamefont {Sahl}},\ and\ \bibinfo {author}
  {\bibfnamefont {S.~W.}\ \bibnamefont {Hell}},\ }\bibfield  {title} {\bibinfo
  {title} {Sted nanoscopy with fluorescent quantum dots},\ }\href
  {https://doi.org/10.1038/ncomms8127} {\bibfield  {journal} {\bibinfo
  {journal} {Nature communications}\ }\textbf {\bibinfo {volume} {6}},\
  \bibinfo {pages} {7127} (\bibinfo {year} {2015})}\BibitemShut {NoStop}%
\bibitem [{\citenamefont {Bruschini}\ \emph {et~al.}(2019)\citenamefont
  {Bruschini}, \citenamefont {Homulle}, \citenamefont {Antolovic},
  \citenamefont {Burri},\ and\ \citenamefont {Charbon}}]{bruschini2019single}%
  \BibitemOpen
  \bibfield  {author} {\bibinfo {author} {\bibfnamefont {C.}~\bibnamefont
  {Bruschini}}, \bibinfo {author} {\bibfnamefont {H.}~\bibnamefont {Homulle}},
  \bibinfo {author} {\bibfnamefont {I.~M.}\ \bibnamefont {Antolovic}}, \bibinfo
  {author} {\bibfnamefont {S.}~\bibnamefont {Burri}},\ and\ \bibinfo {author}
  {\bibfnamefont {E.}~\bibnamefont {Charbon}},\ }\bibfield  {title} {\bibinfo
  {title} {Single-photon avalanche diode imagers in biophotonics: review and
  outlook},\ }\href {https://doi.org/10.1038/s41377-019-0191-5} {\bibfield
  {journal} {\bibinfo  {journal} {Light: Science \& Applications}\ }\textbf
  {\bibinfo {volume} {8}},\ \bibinfo {pages} {87} (\bibinfo {year}
  {2019})}\BibitemShut {NoStop}%
\bibitem [{\citenamefont {Urban}\ \emph {et~al.}(2021)\citenamefont {Urban},
  \citenamefont {Chiang}, \citenamefont {Hammond}, \citenamefont {Cogan},
  \citenamefont {Litzburg}, \citenamefont {Burke}, \citenamefont {Stern},
  \citenamefont {Gelbard}, \citenamefont {Nilsson},\ and\ \citenamefont
  {Krauss}}]{urban2021quantum}%
  \BibitemOpen
  \bibfield  {author} {\bibinfo {author} {\bibfnamefont {J.~M.}\ \bibnamefont
  {Urban}}, \bibinfo {author} {\bibfnamefont {W.}~\bibnamefont {Chiang}},
  \bibinfo {author} {\bibfnamefont {J.~W.}\ \bibnamefont {Hammond}}, \bibinfo
  {author} {\bibfnamefont {N.~M.}\ \bibnamefont {Cogan}}, \bibinfo {author}
  {\bibfnamefont {A.}~\bibnamefont {Litzburg}}, \bibinfo {author}
  {\bibfnamefont {R.}~\bibnamefont {Burke}}, \bibinfo {author} {\bibfnamefont
  {H.~A.}\ \bibnamefont {Stern}}, \bibinfo {author} {\bibfnamefont {H.~A.}\
  \bibnamefont {Gelbard}}, \bibinfo {author} {\bibfnamefont {B.~L.}\
  \bibnamefont {Nilsson}},\ and\ \bibinfo {author} {\bibfnamefont {T.~D.}\
  \bibnamefont {Krauss}},\ }\bibfield  {title} {\bibinfo {title} {Quantum dots
  for improved single-molecule localization microscopy},\ }\href
  {https://doi.org/10.1021/acs.jpcb.0c11545} {\bibfield  {journal} {\bibinfo
  {journal} {The Journal of Physical Chemistry B}\ }\textbf {\bibinfo {volume}
  {125}},\ \bibinfo {pages} {2566} (\bibinfo {year} {2021})}\BibitemShut
  {NoStop}%
\bibitem [{\citenamefont {Triggiani}\ and\ \citenamefont
  {Tamma}(2024)}]{triggiani2024estimation}%
  \BibitemOpen
  \bibfield  {author} {\bibinfo {author} {\bibfnamefont {D.}~\bibnamefont
  {Triggiani}}\ and\ \bibinfo {author} {\bibfnamefont {V.}~\bibnamefont
  {Tamma}},\ }\bibfield  {title} {\bibinfo {title} {Estimation with ultimate
  quantum precision of the transverse displacement between two photons via
  two-photon interference sampling measurements},\ }\href
  {https://doi.org/10.1103/PhysRevLett.132.180802} {\bibfield  {journal}
  {\bibinfo  {journal} {Phys. Rev. Lett.}\ }\textbf {\bibinfo {volume} {132}},\
  \bibinfo {pages} {180802} (\bibinfo {year} {2024})}\BibitemShut {NoStop}%
\bibitem [{\citenamefont {Helstrom}(1969)}]{helstrom1969quantum}%
  \BibitemOpen
  \bibfield  {author} {\bibinfo {author} {\bibfnamefont {C.~W.}\ \bibnamefont
  {Helstrom}},\ }\bibfield  {title} {\bibinfo {title} {Quantum detection and
  estimation theory},\ }\href {https://doi.org/10.1007/BF01007479} {\bibfield
  {journal} {\bibinfo  {journal} {Journal of Statistical Physics}\ }\textbf
  {\bibinfo {volume} {1}},\ \bibinfo {pages} {231} (\bibinfo {year}
  {1969})}\BibitemShut {NoStop}%
\bibitem [{\citenamefont {Fujiwara}\ and\ \citenamefont
  {Nagaoka}(1995)}]{FUJIWARA1995119}%
  \BibitemOpen
  \bibfield  {author} {\bibinfo {author} {\bibfnamefont {A.}~\bibnamefont
  {Fujiwara}}\ and\ \bibinfo {author} {\bibfnamefont {H.}~\bibnamefont
  {Nagaoka}},\ }\bibfield  {title} {\bibinfo {title} {Quantum fisher metric and
  estimation for pure state models},\ }\href
  {https://doi.org/https://doi.org/10.1016/0375-9601(95)00269-9} {\bibfield
  {journal} {\bibinfo  {journal} {Physics Letters A}\ }\textbf {\bibinfo
  {volume} {201}},\ \bibinfo {pages} {119} (\bibinfo {year}
  {1995})}\BibitemShut {NoStop}%
\bibitem [{\citenamefont {Laibacher}\ and\ \citenamefont
  {Tamma}(2015)}]{laibacher2015from}%
  \BibitemOpen
  \bibfield  {author} {\bibinfo {author} {\bibfnamefont {S.}~\bibnamefont
  {Laibacher}}\ and\ \bibinfo {author} {\bibfnamefont {V.}~\bibnamefont
  {Tamma}},\ }\bibfield  {title} {\bibinfo {title} {From the physics to the
  computational complexity of multiboson correlation interference},\ }\href
  {https://doi.org/10.1103/PhysRevLett.115.243605} {\bibfield  {journal}
  {\bibinfo  {journal} {Phys. Rev. Lett.}\ }\textbf {\bibinfo {volume} {115}},\
  \bibinfo {pages} {243605} (\bibinfo {year} {2015})}\BibitemShut {NoStop}%
\bibitem [{\citenamefont {Tamma}\ and\ \citenamefont
  {Laibacher}(2016)}]{tamma2016multi}%
  \BibitemOpen
  \bibfield  {author} {\bibinfo {author} {\bibfnamefont {V.}~\bibnamefont
  {Tamma}}\ and\ \bibinfo {author} {\bibfnamefont {S.}~\bibnamefont
  {Laibacher}},\ }\bibfield  {title} {\bibinfo {title} {Multi-boson correlation
  sampling},\ }\href {https://doi.org/10.1007/s11128-015-1177-8} {\bibfield
  {journal} {\bibinfo  {journal} {Quantum Information Processing}\ }\textbf
  {\bibinfo {volume} {15}},\ \bibinfo {pages} {1241} (\bibinfo {year}
  {2016})}\BibitemShut {NoStop}%
\bibitem [{\citenamefont {Ober}\ \emph {et~al.}(2004)\citenamefont {Ober},
  \citenamefont {Ram},\ and\ \citenamefont {Ward}}]{ober2004localization}%
  \BibitemOpen
  \bibfield  {author} {\bibinfo {author} {\bibfnamefont {R.~J.}\ \bibnamefont
  {Ober}}, \bibinfo {author} {\bibfnamefont {S.}~\bibnamefont {Ram}},\ and\
  \bibinfo {author} {\bibfnamefont {E.~S.}\ \bibnamefont {Ward}},\ }\bibfield
  {title} {\bibinfo {title} {Localization accuracy in single-molecule
  microscopy},\ }\href {https://doi.org/10.1016/S0006-3495(04)74193-4}
  {\bibfield  {journal} {\bibinfo  {journal} {Biophysical journal}\ }\textbf
  {\bibinfo {volume} {86}},\ \bibinfo {pages} {1185} (\bibinfo {year}
  {2004})}\BibitemShut {NoStop}%
\bibitem [{\citenamefont {Lelek}\ \emph {et~al.}(2021)\citenamefont {Lelek},
  \citenamefont {Gyparaki}, \citenamefont {Beliu}, \citenamefont {Schueder},
  \citenamefont {Griffi{\'e}}, \citenamefont {Manley}, \citenamefont
  {Jungmann}, \citenamefont {Sauer}, \citenamefont {Lakadamyali},\ and\
  \citenamefont {Zimmer}}]{lelek2021single}%
  \BibitemOpen
  \bibfield  {author} {\bibinfo {author} {\bibfnamefont {M.}~\bibnamefont
  {Lelek}}, \bibinfo {author} {\bibfnamefont {M.~T.}\ \bibnamefont {Gyparaki}},
  \bibinfo {author} {\bibfnamefont {G.}~\bibnamefont {Beliu}}, \bibinfo
  {author} {\bibfnamefont {F.}~\bibnamefont {Schueder}}, \bibinfo {author}
  {\bibfnamefont {J.}~\bibnamefont {Griffi{\'e}}}, \bibinfo {author}
  {\bibfnamefont {S.}~\bibnamefont {Manley}}, \bibinfo {author} {\bibfnamefont
  {R.}~\bibnamefont {Jungmann}}, \bibinfo {author} {\bibfnamefont
  {M.}~\bibnamefont {Sauer}}, \bibinfo {author} {\bibfnamefont
  {M.}~\bibnamefont {Lakadamyali}},\ and\ \bibinfo {author} {\bibfnamefont
  {C.}~\bibnamefont {Zimmer}},\ }\bibfield  {title} {\bibinfo {title}
  {Single-molecule localization microscopy},\ }\href
  {https://doi.org/10.1038/s43586-021-00038-x} {\bibfield  {journal} {\bibinfo
  {journal} {Nature reviews methods primers}\ }\textbf {\bibinfo {volume}
  {1}},\ \bibinfo {pages} {39} (\bibinfo {year} {2021})}\BibitemShut {NoStop}%
\bibitem [{\citenamefont {Salazar-Serrano}\ \emph {et~al.}(2015)\citenamefont
  {Salazar-Serrano}, \citenamefont {Valencia},\ and\ \citenamefont
  {Torres}}]{salazar2015tunable}%
  \BibitemOpen
  \bibfield  {author} {\bibinfo {author} {\bibfnamefont {L.~J.}\ \bibnamefont
  {Salazar-Serrano}}, \bibinfo {author} {\bibfnamefont {A.}~\bibnamefont
  {Valencia}},\ and\ \bibinfo {author} {\bibfnamefont {J.~P.}\ \bibnamefont
  {Torres}},\ }\bibfield  {title} {\bibinfo {title} {Tunable beam displacer},\
  }\bibfield  {journal} {\bibinfo  {journal} {Review of Scientific
  Instruments}\ }\textbf {\bibinfo {volume} {86}},\ \href
  {https://doi.org/10.1063/1.4914834} {10.1063/1.4914834} (\bibinfo {year}
  {2015})\BibitemShut {NoStop}%
\bibitem [{\citenamefont {Pirandola}\ \emph {et~al.}(2018)\citenamefont
  {Pirandola}, \citenamefont {Bardhan}, \citenamefont {Gehring}, \citenamefont
  {Weedbrook},\ and\ \citenamefont {Lloyd}}]{pirandola2018advances}%
  \BibitemOpen
  \bibfield  {author} {\bibinfo {author} {\bibfnamefont {S.}~\bibnamefont
  {Pirandola}}, \bibinfo {author} {\bibfnamefont {B.~R.}\ \bibnamefont
  {Bardhan}}, \bibinfo {author} {\bibfnamefont {T.}~\bibnamefont {Gehring}},
  \bibinfo {author} {\bibfnamefont {C.}~\bibnamefont {Weedbrook}},\ and\
  \bibinfo {author} {\bibfnamefont {S.}~\bibnamefont {Lloyd}},\ }\bibfield
  {title} {\bibinfo {title} {Advances in photonic quantum sensing},\ }\href
  {https://doi.org/10.1038/s41566-018-0301-6} {\bibfield  {journal} {\bibinfo
  {journal} {Nature Photonics}\ }\textbf {\bibinfo {volume} {12}},\ \bibinfo
  {pages} {724} (\bibinfo {year} {2018})}\BibitemShut {NoStop}%
\bibitem [{\citenamefont {Kim}\ \emph {et~al.}(2006{\natexlab{b}})\citenamefont
  {Kim}, \citenamefont {Fiorentino},\ and\ \citenamefont
  {Wong}}]{kim2006phase}%
  \BibitemOpen
  \bibfield  {author} {\bibinfo {author} {\bibfnamefont {T.}~\bibnamefont
  {Kim}}, \bibinfo {author} {\bibfnamefont {M.}~\bibnamefont {Fiorentino}},\
  and\ \bibinfo {author} {\bibfnamefont {F.~N.~C.}\ \bibnamefont {Wong}},\
  }\bibfield  {title} {\bibinfo {title} {Phase-stable source of
  polarization-entangled photons using a polarization sagnac interferometer},\
  }\href {https://doi.org/10.1103/PhysRevA.73.012316} {\bibfield  {journal}
  {\bibinfo  {journal} {Phys. Rev. A}\ }\textbf {\bibinfo {volume} {73}},\
  \bibinfo {pages} {012316} (\bibinfo {year} {2006}{\natexlab{b}})}\BibitemShut
  {NoStop}%
\bibitem [{\citenamefont {Chen}\ \emph {et~al.}(2018)\citenamefont {Chen},
  \citenamefont {Ecker}, \citenamefont {Wengerowsky}, \citenamefont {Bulla},
  \citenamefont {Joshi}, \citenamefont {Steinlechner},\ and\ \citenamefont
  {Ursin}}]{2018chenpolarization}%
  \BibitemOpen
  \bibfield  {author} {\bibinfo {author} {\bibfnamefont {Y.}~\bibnamefont
  {Chen}}, \bibinfo {author} {\bibfnamefont {S.}~\bibnamefont {Ecker}},
  \bibinfo {author} {\bibfnamefont {S.}~\bibnamefont {Wengerowsky}}, \bibinfo
  {author} {\bibfnamefont {L.}~\bibnamefont {Bulla}}, \bibinfo {author}
  {\bibfnamefont {S.~K.}\ \bibnamefont {Joshi}}, \bibinfo {author}
  {\bibfnamefont {F.}~\bibnamefont {Steinlechner}},\ and\ \bibinfo {author}
  {\bibfnamefont {R.}~\bibnamefont {Ursin}},\ }\bibfield  {title} {\bibinfo
  {title} {Polarization entanglement by time-reversed hong-ou-mandel
  interference},\ }\href {https://doi.org/10.1103/PhysRevLett.121.200502}
  {\bibfield  {journal} {\bibinfo  {journal} {Phys. Rev. Lett.}\ }\textbf
  {\bibinfo {volume} {121}},\ \bibinfo {pages} {200502} (\bibinfo {year}
  {2018})}\BibitemShut {NoStop}%
\bibitem [{\citenamefont {Pang}\ \emph {et~al.}(2008)\citenamefont {Pang},
  \citenamefont {Beckham},\ and\ \citenamefont {Meissner}}]{pang2008quantum}%
  \BibitemOpen
  \bibfield  {author} {\bibinfo {author} {\bibfnamefont {S.}~\bibnamefont
  {Pang}}, \bibinfo {author} {\bibfnamefont {R.~E.}\ \bibnamefont {Beckham}},\
  and\ \bibinfo {author} {\bibfnamefont {K.~E.}\ \bibnamefont {Meissner}},\
  }\bibfield  {title} {\bibinfo {title} {Quantum dot-embedded microspheres for
  remote refractive index sensing},\ }\bibfield  {journal} {\bibinfo  {journal}
  {Applied physics letters}\ }\textbf {\bibinfo {volume} {92}},\ \href
  {https://doi.org/10.1063/1.2937209} {10.1063/1.2937209} (\bibinfo {year}
  {2008})\BibitemShut {NoStop}%
\bibitem [{\citenamefont {Serbyn}\ \emph {et~al.}(2014)\citenamefont {Serbyn},
  \citenamefont {Knap}, \citenamefont {Gopalakrishnan}, \citenamefont
  {Papi\ifmmode~\acute{c}\else \'{c}\fi{}}, \citenamefont {Yao}, \citenamefont
  {Laumann}, \citenamefont {Abanin}, \citenamefont {Lukin},\ and\ \citenamefont
  {Demler}}]{serbyn2014interferometric}%
  \BibitemOpen
  \bibfield  {author} {\bibinfo {author} {\bibfnamefont {M.}~\bibnamefont
  {Serbyn}}, \bibinfo {author} {\bibfnamefont {M.}~\bibnamefont {Knap}},
  \bibinfo {author} {\bibfnamefont {S.}~\bibnamefont {Gopalakrishnan}},
  \bibinfo {author} {\bibfnamefont {Z.}~\bibnamefont
  {Papi\ifmmode~\acute{c}\else \'{c}\fi{}}}, \bibinfo {author} {\bibfnamefont
  {N.~Y.}\ \bibnamefont {Yao}}, \bibinfo {author} {\bibfnamefont {C.~R.}\
  \bibnamefont {Laumann}}, \bibinfo {author} {\bibfnamefont {D.~A.}\
  \bibnamefont {Abanin}}, \bibinfo {author} {\bibfnamefont {M.~D.}\
  \bibnamefont {Lukin}},\ and\ \bibinfo {author} {\bibfnamefont {E.~A.}\
  \bibnamefont {Demler}},\ }\bibfield  {title} {\bibinfo {title}
  {Interferometric probes of many-body localization},\ }\href
  {https://doi.org/10.1103/PhysRevLett.113.147204} {\bibfield  {journal}
  {\bibinfo  {journal} {Phys. Rev. Lett.}\ }\textbf {\bibinfo {volume} {113}},\
  \bibinfo {pages} {147204} (\bibinfo {year} {2014})}\BibitemShut {NoStop}%
\bibitem [{\citenamefont {Barer}(1957)}]{barer1957refractometry}%
  \BibitemOpen
  \bibfield  {author} {\bibinfo {author} {\bibfnamefont {R.}~\bibnamefont
  {Barer}},\ }\bibfield  {title} {\bibinfo {title} {Refractometry and
  interferometry of living cells},\ }\href
  {https://doi.org/10.1364/JOSA.47.000545} {\bibfield  {journal} {\bibinfo
  {journal} {Journal of the Optical Society of America}\ }\textbf {\bibinfo
  {volume} {47}},\ \bibinfo {pages} {545} (\bibinfo {year} {1957})}\BibitemShut
  {NoStop}%
\bibitem [{\citenamefont {Persson}\ \emph {et~al.}(2003)\citenamefont
  {Persson}, \citenamefont {Yoshida}, \citenamefont {Tong}, \citenamefont
  {Reinhold},\ and\ \citenamefont {Burgd\"orfer}}]{persson2003quantum}%
  \BibitemOpen
  \bibfield  {author} {\bibinfo {author} {\bibfnamefont {E.}~\bibnamefont
  {Persson}}, \bibinfo {author} {\bibfnamefont {S.}~\bibnamefont {Yoshida}},
  \bibinfo {author} {\bibfnamefont {X.-M.}\ \bibnamefont {Tong}}, \bibinfo
  {author} {\bibfnamefont {C.~O.}\ \bibnamefont {Reinhold}},\ and\ \bibinfo
  {author} {\bibfnamefont {J.}~\bibnamefont {Burgd\"orfer}},\ }\bibfield
  {title} {\bibinfo {title} {Quantum localization in the three-dimensional
  kicked rydberg atom},\ }\href {https://doi.org/10.1103/PhysRevA.68.063406}
  {\bibfield  {journal} {\bibinfo  {journal} {Phys. Rev. A}\ }\textbf {\bibinfo
  {volume} {68}},\ \bibinfo {pages} {063406} (\bibinfo {year}
  {2003})}\BibitemShut {NoStop}%
\bibitem [{\citenamefont {Wang}\ \emph {et~al.}(2021)\citenamefont {Wang},
  \citenamefont {Xu}, \citenamefont {chi Li},\ and\ \citenamefont
  {Zhang}}]{Wang2021quantum}%
  \BibitemOpen
  \bibfield  {author} {\bibinfo {author} {\bibfnamefont {B.}~\bibnamefont
  {Wang}}, \bibinfo {author} {\bibfnamefont {L.}~\bibnamefont {Xu}}, \bibinfo
  {author} {\bibfnamefont {J.}~\bibnamefont {chi Li}},\ and\ \bibinfo {author}
  {\bibfnamefont {L.}~\bibnamefont {Zhang}},\ }\bibfield  {title} {\bibinfo
  {title} {Quantum-limited localization and resolution in three dimensions},\
  }\href {https://doi.org/10.1364/PRJ.417613} {\bibfield  {journal} {\bibinfo
  {journal} {Photon. Res.}\ }\textbf {\bibinfo {volume} {9}},\ \bibinfo {pages}
  {1522} (\bibinfo {year} {2021})}\BibitemShut {NoStop}%
\bibitem [{\citenamefont {Aharonovich}\ \emph {et~al.}(2016)\citenamefont
  {Aharonovich}, \citenamefont {Englund},\ and\ \citenamefont
  {Toth}}]{aharonovich2016solid}%
  \BibitemOpen
  \bibfield  {author} {\bibinfo {author} {\bibfnamefont {I.}~\bibnamefont
  {Aharonovich}}, \bibinfo {author} {\bibfnamefont {D.}~\bibnamefont
  {Englund}},\ and\ \bibinfo {author} {\bibfnamefont {M.}~\bibnamefont
  {Toth}},\ }\bibfield  {title} {\bibinfo {title} {Solid-state single-photon
  emitters},\ }\href {https://doi.org/10.1038/nphoton.2016.186} {\bibfield
  {journal} {\bibinfo  {journal} {Nature photonics}\ }\textbf {\bibinfo
  {volume} {10}},\ \bibinfo {pages} {631} (\bibinfo {year} {2016})}\BibitemShut
  {NoStop}%
\end{thebibliography}%

\newpage
\appendix
\begin{widetext}
\newpage
\section{Supplemental Material}
\subsection{Quantum mechanical derivation of the HOM effect}
The two-photon state used in our spatial HOM interferometry is generated from a spontaneous parametric down conversion (SPDC) process. With the assistance of spatial filtering that is implemented by coupling down converted photons into single mode fiber, the resultant two-photon state can be described as
\begin{equation}
	\ket{\Psi}=\int d x_1\psi_1(x_1)\hat{a}_1^\dagger(x_1) \ket{0}\otimes\int d x_2\psi_2(x_2)\hat{a}_2^\dagger(x_2) \ket{0}.
\end{equation}
In order to fulfill the task of parameter sensing, we use this probe state to interact with dynamic system, i.e., introduce a relative transverse deflection $\Delta \theta$ in the path of first photon. It implies that a relative phase shift of $\exp(-ik_1\Delta \theta d)$ is added, and transforming the state into
\begin{equation}
\ket{\psi}=\int_{-\infty}^\infty\int_{-\infty}^\infty dx_1dx_2e^{-ik_1\Delta \theta d}\psi_1(x_1)\psi_2(x_2)\hat{a}_1^\dag(x_1)\hat{a}_2^\dag(x_2)\ket{0}.
\end{equation}
Then we apply the operation of a balanced beam splitter to transform this state as
\begin{equation}
\begin{split}
\hat{a}_1^\dag(x_1)=\frac{1}{\sqrt{2}}[\hat{a}_3^\dag(x_1)+\hat{a}_4^\dag(x_1)]\\
\hat{a}_2^\dag(x_2)=\frac{1}{\sqrt{2}}[\hat{a}_3^\dag(x_2)-\hat{a}_4^\dag(x_2)],\\
\end{split}
\end{equation}
where subscripts 1/2 (3/4) represent two input (output) ports of the beam splitter. Thus we get the transformed state as
\begin{equation}\label{eq:after bs}
\begin{split}
\ket{\psi}\rightarrow&\frac{1}{2}\int_{-\infty}^\infty\int_{-\infty}^\infty dx_1dx_2\psi_1(x_1)\psi_2(x_2)e^{-ik_1\Delta\theta d }[\hat{a}_3^\dag(x_1)\hat{a}_3^\dag(x_2)-\hat{a}_4^\dag(x_1)\hat{a}_4^\dag(x_2)\\
&+\hat{a}_3^\dag(x_2)\hat{a}_4^\dag(x_1)-\hat{a}_3^\dag(x_1)\hat{a}_4^\dag(x_2)]\ket{0}.
\end{split}
\end{equation}
As a direct consequence of post-selection by registering the coincidence events in two distinct spatial modes, only last two terms of \eqref{eq:after bs} can survive. So it can be simplified to
\begin{equation}
\begin{split}
\ket{\psi_A}&=\frac{1}{2}\int_{-\infty}^\infty\int_{-\infty}^\infty dx_1dx_2\psi_1(x_1)\psi_2(x_2)e^{-ik_1\Delta \theta d}[\hat{a}_3^\dag(x_2)\hat{a}_4^\dag(x_1)-\hat{a}_3^\dag(x_1)\hat{a}_4^\dag(x_2)]\ket{0}
\end{split}
\end{equation}
Since the paired photons are symmetric when they are exchanged, we can simplify the above inequality as
\begin{equation}
\ket{\psi_A}=\frac{1}{2}\int_{-\infty}^\infty\int_{-\infty}^\infty dx_1dx_2\psi_1(x_1)\psi_2(x_2)[1-e^{-i(k_1-k_2)\Delta\theta d}]\hat{a}_3^\dag(x_1)\hat{a}_4^\dag(x_2)\ket{0}.\\
\end{equation}
The detection operators of two detectors in different output modes are
\begin{equation}
\begin{split}
\hat{E}_3^{(+)}=\frac{1}{\sqrt{2\pi}}\int_{-\infty}^\infty dx_1\hat{a}_3(x_1)e^{-ik_1x_1},\\
\hat{E}_4^{(+)}=\frac{1}{\sqrt{2\pi}}\int_{-\infty}^\infty dx_2\hat{a}_4(x_2)e^{-ik_2x_2}.\\
\end{split}
\end{equation}
Thus we can calculate $\hat{E}_4^{(+)}\hat{E}_3^{(+)}\ket{\psi_A}$ as
\begin{equation}
\begin{split}
\hat{E}_4^{(+)}\hat{E}_3^{(+)}\ket{\psi_A}=&\frac{1}{2\pi}\int_{-\infty}^\infty\int_{-\infty}^\infty dx_1dx_2\hat{a}_3(x_1)\hat{a}_4(x_2)e^{-ik_1x_1 }e^{-ik_2x_2}\\
&\times\frac{1}{2}\int_{-\infty}^\infty\int_{-\infty}^\infty dx_1^\prime dx_2^\prime \psi_1(x_1^\prime)\psi_2(x_2^\prime)[1-e^{-i(k_1^\prime-k_2^\prime)\Delta\theta d}]\hat{a}_3^\dag(x_1^\prime)\hat{a}_4^\dag(x_2^\prime)\ket{0}\\
=&\frac{1}{4\pi}\int_{-\infty}^\infty\int_{-\infty}^\infty dx_1dx_2\psi_1(x_1)\psi_2(x_2)[1-e^{-i(k_1^\prime-k_2^\prime)\Delta\theta d}]e^{-ik_1x_1}e^{-ik_2x_2},
\end{split}
\end{equation}
where we add $k^\prime$ to distinguish the symbols from detection or photons, albeit $k^\prime=k$. Finally the coincidence probability $P(\Delta \theta)$ as a function of time delay can be expressed as
\begin{equation}
\begin{split}
P(\Delta \theta)=&\langle\psi_A|\hat{E}_3^{(-)}\hat{E}_4^{(-)}\hat{E}_4^{(+)}\hat{E}_3^{(+)}\ket{\psi_A}\\
=&\frac{1}{4}|\varphi(k_1)|^2|\varphi(k_2)|^2(1-e^{-i(k_1-k_2)\Delta\theta d})(1-e^{i(k_1-k_2)\Delta\theta d})\\
=&\frac{1}{2}|\varphi(k_1)|^2|\varphi(k_2)|^2[1-\cos((k_1-k_2)\Delta \theta d)]
\end{split}
\end{equation}
When the transverse deflection $\Delta \theta=0$ that is set in a conventional HOM interferometer, $P(\Delta \theta)$ exists a well-known dip.

For a concrete example, we consider the paradigmatic case of a Gaussian spatial function to obtain the normalized probability of coincidence counts as a function of relative transverse deflection $\Delta \theta$ as
\begin{equation}\label{coincidence probability}
\begin{split}
P_c(\Delta k, \Delta \theta)=\frac{1}{2}C(\Delta k)[1-\cos(\Delta k\Delta \theta d)],
\end{split}
\end{equation}
where $C(\Delta k)$ represents the transverse momentum distribution, $\Delta k=|k_1-k_2|$ is the relative transverse momentum.
\subsection{Loss model}
Since real experiments suffer from losses, dominated by transmission loss and inefficient photon detectors, we therefore have a model by allowing for a photon to be lost with probability $\gamma$. The full mode is given by
\begin{equation}
\left(
\begin{array}{c}
P_0\\
P_1\\
P_2\\
\end{array}
\right)
=
\left(
\begin{array}{cc}
\gamma^2&\gamma^2\\
2\gamma(1-\gamma)&1-\gamma^2\\
1-2\gamma(1-\gamma)-\gamma^2&0\\
\end{array}
\right)
\left(
\begin{array}{c}
P_c(\tau)\\
P_b(\tau)\\
\end{array}
\right)
\end{equation}
with $P_b(\tau)=1-P_c(\tau)$ implied by normalization.

\subsection{Cramér-Rao bound and sensitivity}
\subsubsection{Fundamentals}
Assisted by quantum information, we use a pure state to estimate the unknown value of a variable. In our case, the variable is a transverse deflection $\Delta \theta$. For the sake of simplicity, we substitute $\Delta \theta$ with $\theta$. After interaction with the system that produces the delay, the pure state is modified and can be written as $\ket{\Psi(\theta)}$. We measure something in a given measurement using $\ket{\Psi(\theta)}$, from where we estimate the value of $\theta$. The precision of the estimator $\delta \theta$ will always be:
\begin{equation}
\delta\theta\geq\frac{1}{2Q^{1/2}}=\delta\theta_{CR},
\end{equation}
where
\begin{equation}
Q=\langle\frac{\partial\Psi(\theta)}{\partial\theta}\ket{\frac{\partial\Psi(\theta)}{\partial\theta}}-|\langle\Psi(\theta)\ket{\frac{\partial\Psi(\theta)}{\partial\theta}}|^2.
\end{equation}
This is the quantum Cramér-Rao bound, which is the ultimate limit of sensitivity that can be achieved. Any experiment that we could perform cannot provide a better result than this.

\subsubsection{The Cramér-Rao bound}
We consider a pair of indistinguishable photons, whose state of interest can be written as
\begin{equation}
\ket{\Psi(\theta)}=\int d\Omega f(\Omega)\exp[i(k_1^0+\Omega)\theta d]a_1^\dag(k_1^0+\Omega)a_2^\dag(k_2^0)\ket{vac},
\end{equation}
where $\Omega$ represents the Gaussian spatial amplitude function with $\int d\Omega|f(\Omega)|^2=1$ and we assume that the first photon denoted by the subscript $1$ is tilted by a transverse deflection $\theta$.
Let us calculate Q, We obtain
\begin{equation}
\langle\frac{\partial\Psi(\theta)}{\partial\theta}\ket{\frac{\partial\Psi(\theta)}{\partial\theta}}=\int d\Omega|f(\Omega)|^2(k_1^0+\Omega)^2d^2=[(k_1^0)^2+2k_1^0\langle\Omega\rangle+\langle\Omega^2\rangle]d^2,
\end{equation}
and
\begin{equation}
\langle\Psi(\theta)\ket{\frac{\partial\Psi(\theta)}{\partial\theta}}=i\int d\Omega|f(\Omega)|^2(k_1^0+\Omega)d=id(k_1^0+\langle\Omega\rangle).
\end{equation}
Here $\langle\Omega\rangle=\int d\Omega\Omega|f(\Omega)|^2$ and $\langle\Omega^2\rangle=\int d\Omega\Omega^2|f(\Omega)|^2$.
We have
\begin{equation}
Q=[(k_1^0)^2+2k_1^0\langle\Omega\rangle+\langle\Omega^2\rangle]d^2-[d(k_1^0+\langle\Omega\rangle)]^2=(\langle\Omega^2\rangle-\langle\Omega\rangle^2)d^2\equiv2\sigma_k^2d^2,
\end{equation}
$\sigma_k$ is the spatial RMS bandwidth of single photons. Therefore the Cramér-Rao bound is
\begin{equation}
\delta\theta_{CR}=\frac{1}{2\sqrt{2}\sigma_kd}.
\end{equation}
This limit only depends on the spatial bandwidth of single photons, and the distance between the source and detector.

\subsection{Fisher information}
We have calculated the Fisher information used to estimate the transverse deflection for various imperfect visibilities and channel-loss rates by following the statistical model described in the main text. In the context of our experimental setup, the Fisher information is proportional to the standard deviation of the transverse-momentum distribution of single photons and the distance between the source and the detector as shown in Fig. \ref{figS1}.
\begin{figure}[h]
	\centering
	\includegraphics[width=1\linewidth]{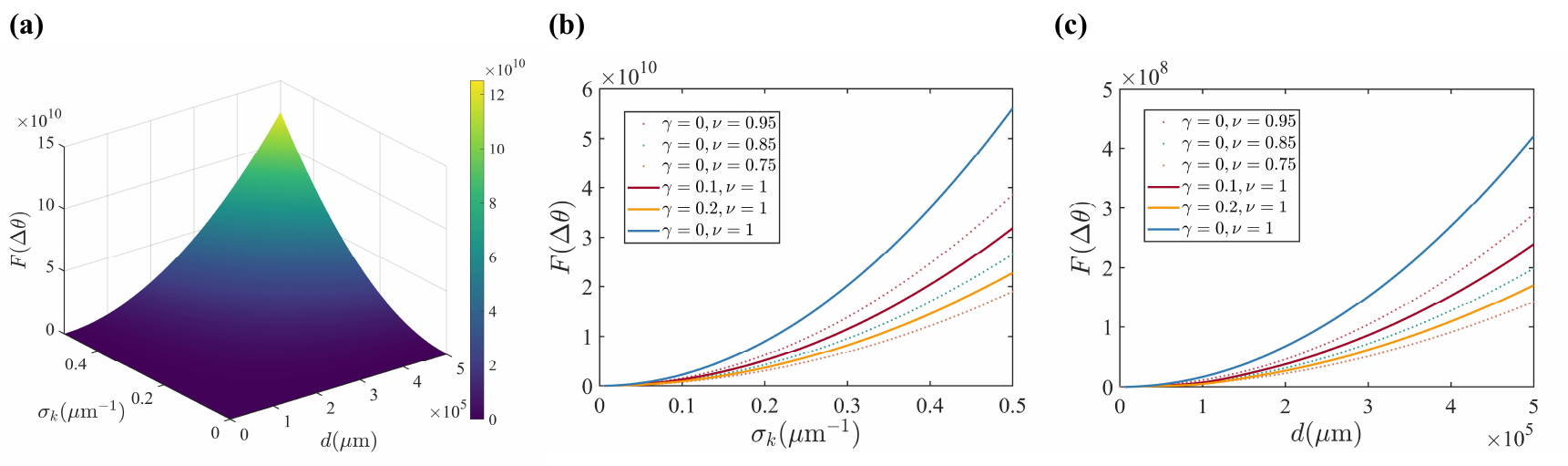}
	\caption{Theoretical prediction of Fisher information for transverse deflection measurement. (a) Fisher information as functions of $d$ and $\sigma_k$ ($\Delta \theta=0.52$ mrad). (b) Fisher information as a function of standard deviation of the transverse-momentum distribution of single photons $\sigma_k$ ($\Delta \theta=0.52$ mrad, $d=335$ mm) and (c) as a function of distance between the sources and detectors $d$ ($\Delta \theta=0.52$ mrad, $\sigma_k$=0.029 $\mu$m$^{-1}$)for various imperfect visibilities and channel-loss rates.}
	\label{figS1}
\end{figure}

%

\end{widetext}
\end{document}